\documentclass[12pt,a4paper]{article}
\usepackage{graphicx}
\usepackage[T1]{fontenc}
\usepackage[utf8]{inputenc}
\usepackage{textcomp}
\usepackage[sc,osf]{mathpazo}
\usepackage{a4wide}  
\usepackage{latexsym,amsthm,amsfonts,amsmath,mathrsfs,amssymb}
\usepackage{dsfont}
\usepackage{accents}
\usepackage[nosort]{cite}
\usepackage{booktabs} 
\usepackage[unicode,implicit]{hyperref}
\hypersetup{%
  pdftitle    = {Non-supersymmetric black holes with $\alpha'$ corrections}
  pdfkeywords = {black hole, string theory, entropy, temperature, Wald, Hawking,
    branes, supergravity, supersymmetry, corrections, WGC, weak gravity
    conjecture, heterotic superstring},
  pdfauthor   = {Pablo A. Cano, Tom\'as Ort\'{\i}n, Alejandro Ruip\'erez and
    Matteo Zatti},
  plainpages  = true,
  colorlinks  = true,
  citecolor   = blue,
  urlcolor    = red,
  linkcolor   = black
}
\newcommand{\hepth}[1]{{\tt
\href{http://www.arXiv.org/abs/hep-th/#1}{hep-th/#1}}}
\newcommand{\grqc}[1]{{\tt
\href{http://www.arXiv.org/abs/gr-qc/#1}{gr-qc/#1}}}

\newcommand{\arxiv}[1]{{\tt arXiv:\href{http://www.arXiv.org/abs/#1}{#1}}}
\newcommand{\doi}[1]{{\tt DOI:\href{http://dx.doi.org/#1}{#1}}}

\makeatletter
\@addtoreset{equation}{section}
\makeatother

\begin{document}

\allowdisplaybreaks

\begin{flushright}
\small
IFT-UAM/CSIC-21-116\\
December 17\textsuperscript{th}, 2021\\
\normalsize
\end{flushright}

\vspace{.5cm}

\begin{center}

{\Large {\bf Non-supersymmetric black holes with $\alpha'$ corrections}}

\vspace{1cm}

{\large {\sl Dedicated to the memory of Mees de Roo}}

\vspace{1cm}

\renewcommand{\thefootnote}{\alph{footnote}}

{\sl\large Pablo A.~Cano,}$^{1,}$\footnote{Email: {\tt pabloantonio.cano[at]kuleuven.be}}
{\sl\large  Tom\'{a}s Ort\'{\i}n,}$^{2,}$\footnote{Email: {\tt tomas.ortin[at]csic.es}}
{\sl\large Alejandro Ruip\'erez,}$^{3,4,}$\footnote{Email: {\tt alejandro.ruiperez[at]pd.infn.it}}
{\sl\large and Matteo Zatti}$^{2,}$\footnote{Email: {\tt matteo.zatti[at]estudiante.uam.es}}

\setcounter{footnote}{0}
\renewcommand{\thefootnote}{\arabic{footnote}}

\vspace{0.8cm}

${}^{1}${\it Instituut voor Theoretische Fysica, KU Leuven\\
	Celestijnenlaan 200D, B-3001 Leuven, Belgium}\\

\vspace{0.4cm}

${}^{2}${\it Instituto de F\'{\i}sica Te\'orica UAM/CSIC\\
C/ Nicol\'as Cabrera, 13--15,  C.U.~Cantoblanco, E-28049 Madrid, Spain}

\vspace{0.4cm}

${}^{3}${\it Dipartimento di Fisica ed Astronomia ``Galileo Galilei'',\\
  Universit\`a di Padova, Via Marzolo 8, 35131 Padova, Italy}

\vspace{0.4cm}

${}^{4}${\it INFN, Sezione di Padova,Via Marzolo 8, 35131 Padova, Italy}

\vspace{0.7cm}


{\bf Abstract}
\end{center}
\begin{quotation} {\small We study extremal 5- and 4-dimensional black hole
    solutions of the Heterotic Superstring effective action at first order in
    $\alpha'$ with, respectively, 3 and 4 charges of arbitrary signs. For a
    particular choice of the relative signs of these charges the solutions are
    supersymmetric, but other signs give rise to non-supersymmetric extremal
    black holes.  Although at zeroth order all these solutions are formally
    identical, we show that the $\alpha'$ corrections are drastically
    different depending on how we break supersymmetry. We provide fully
    analytic $\mathcal{O}(\alpha')$ solutions and we compute their charges,
    masses and entropies and check that they are invariant under
    $\alpha'$-corrected T-duality transformations. The masses of some of these
    black holes are corrected in a complicated way, but we show that the shift
    is always negative, in agreement with the Weak Gravity Conjecture. Our
    formula for the corrected entropy of the four-dimensional black holes
    generalizes previous results obtained from the entropy function method, as
    we consider a more general way in which supersymmetry can be broken.  }
\end{quotation}

\newpage
\pagestyle{plain}

\tableofcontents

\newpage

\section{Introduction}

Black holes provide an excellent ground for testing string theory as a quantum
theory of gravity. The statistical derivation of the Bekenstein-Hawking
entropy of 5-dimensional supersymmetric black holes by Strominger and Vafa
\cite{Strominger:1996sh} remains, to this day, one of the main tests passed by
string theory in this direction. Since then, a great deal of work has been
devoted to push this comparison beyond leading order in the large-charge
expansion. On the microscopic side, this implies that one has to compute
corrections to Cardy's formula \cite{Sen:2007qy, Castro:2008ys}. On the
gravitational side, instead, one has to take into account $\alpha'$
corrections to supergravity, which encode genuine stringy
effects. For this comparison to be
possible and meaningful, several things are needed from the gravitational side:

\begin{enumerate}
\item Knowledge of the string effective action to higher order in $\alpha'$. 
\item Black-hole solutions of the higher-order equations derived from those
  actions.
\item A consistent procedure to compute the analog of the Bekenstein-Hawking
  entropy in higher-order theories.
\item A dictionary relating the parameters of the black-hole solutions (mass,
  angular momenta, charges, moduli) to those of the microscopic system.

\end{enumerate}

Let us comment upon each of these points.

\begin{enumerate}
\item Our knowledge of the higher-order in $\alpha'$ corrections to the string
  effective action is very incomplete as they are only known for the first few
  orders. Among the supersymmetric string theories, only the heterotic has
  first-order corrections and these and the second- and third-order ones are
  also known
  \cite{Gross:1986mw,Metsaev:1986yb,Metsaev:1987zx,Bergshoeff:1989de}. This
  makes this theory especially well-suited for the the computation of $\alpha'$
  corrections to black-hole solutions. Since the complexity of the
  calculations increases rapidly with the power of $\alpha'$, only the
  first-order ones have typically been computed.
  
\item Only a few analytic stringy black hole solutions are known to first
  order in $\alpha'$
  \cite{Campbell:1991kz,Natsuume:1994hd,Cano:2018qev, Chimento:2018kop, Cano:2018brq,Cano:2019ycn,
    Cano:2021rey}. Many results on $\alpha'$ corrections to black-hole
  entropies are based on the entropy function formalism \cite{Sen:2005wa}. Its
  main advantage is that only deals with the near-horizon geometries, circumventing the task of solving the corrected equations of motion. It has
  however some limitations because one loses information about the asymptotic
  region, and, for instance, one cannot get the mass of the black holes. Also,
  the interpretation of the charges is arguably less transparent in this
  formalism.

  One of the main goals of this paper is to increase the number of complete
  first-order in $\alpha'$ black hole solutions known by determining the
  first-order $\alpha'$ corrections to all the most basic 3-charge,
  5-dimensional and 4-charge, 4-dimensional extremal black holes. This implies
  allowing for all the possible signs of the charges. Only some combinations
  of the signs of the charges are compatible with supersymmetry, as it is well
  known \cite{Ortin:1996bz}. Some non-supersymmetric choices have been
  considered in the literature \cite{Sen:2007qy,Sahoo:2006pm,Prester:2008iu}
  but, here, we consider all of them and, as we are going to see, the
  $\alpha'$ corrections to the solutions turn out to be wildly different for
  different sign choices.
  
  Besides, these solutions are interesting to test the so-called mild form of
  the Weak Gravity Conjecture
  \cite{Kats:2006xp,Cheung:2018cwt,Hamada:2018dde}, according to which, the
  corrections to the mass of extremal black holes should be
  negative in a consistent theory of quantum gravity. This conjecture has been
  applied in a number of scenarios to constrain the parameters of effective
  field theories
  \cite{Aalsma:2019ryi,Bellazzini:2019xts,Cremonini:2019wdk,Charles:2019qqt,Loges:2019jzs,Andriolo:2020lul,Loges:2020trf,Cano:2020qhy,Cano:2021tfs,Arkani-Hamed:2021ajd}. However,
  by studying the effect of higher-derivative corrections coming explicitly
  from string theory, one can in turn test this conjecture
  \cite{Cano:2019oma,Cano:2019ycn}, as in that case it is expected to be
  satisfied.

\item Using the formalism developed in Ref.~\cite{Lee:1990nz}, Wald showed
  that, in a diffeomorphism-invariant theory of gravity, the black-hole entropy
  can be identified with the Noether charge associated to diffeomorphisms
  evaluated for the Killing vector normal on the event horizon, because this is
  the quantity whose variation appears multiplied by the temperature in the
  first law of black-hole mechanics \cite{Wald:1993nt}. This entropy
  (\textit{Wald entropy}) coincides with the Bekenstein-Hawking entropy for
  General Relativity but can also be used in theories of higher order in the
  curvature such as the higher order in $\alpha'$ string effective
  action. However, this action also contains matter fields, which were not
  considered in Ref.~\cite{Wald:1993nt}.
  
  In Ref.~\cite{Iyer:1994ys} Iyer and Wald considered matter fields, assuming
  they behave as tensors under diffeomorphisms and gave a recipe for computing
  the Wald entropy (the Iyer-Wald entropy formula) which has become standard
  in the literature. However, as discussed in more detail in
  Refs.~\cite{Elgood:2020svt,Elgood:2020mdx,Elgood:2020nls}, the only field in
  the Standard Model that transforms as a tensor is the metric. The rest have
  gauge freedoms and their transformations under diffeomorphisms always
  involve ``compensating gauge transformations'' that have to be properly
  taken into account. Only then one obtains the work terms in the first law of
  black-hole mechanics which, actually, were not derived in
  Ref.~\cite{Iyer:1994ys}.\footnote{Actually, the problem is much worse: as
    observed in \cite{Elgood:2020xwu,Ortin:2020xdm}, the naive application of
    the Iyer-Wald formula to the first-order in $\alpha'$ Heterotic
    Superstring effective action gives a gauge-dependent result: the entropy
    computed in that form depends on the Lorentz frame used. This is, clearly,
    unacceptable. On the other hand, as discussed in Ref.~\cite{Faedo:2019xii}
    if one ignores this problem and makes the calculation in the simplest
    frame, the result seems to differ from the one computed by other methods
    such as Sen's \cite{Sahoo:2006pm}.} These terms arise from additional
  contributions to the diffeomorphisms Noether charge, which means that, in
  general, the entropy is not the Noether charge and the derivation of the
  first law with all the terms that contribute to it is a \textit{conditio
    sine qua non} for the identification of the Wald entropy.

  In Ref.~\cite{Elgood:2020nls} a gauge-invariant entropy formula for the
  Heterotic Superstring effective action at first order in $\alpha'$ was
  derived after taking into account all the complicated gauge transformations
  of the theory. This formula gives unambiguous results and, in the few
  examples in which the entropy has been computed by other methods, it gives
  the same result.

  The second main goal of this paper is to compute the entropy of the
  $\alpha'$-corrected solutions we are going to find using this formula.

\item Finding a stringy interpretation of the parameters occurring in the
  simplest leading-order black-hole solutions presents no special problems
  because, first of all, the Chern-Simons terms that occur in the 4- and
  5-dimensional field strengths vanish and all the definitions of conserved
  charge (Page, Maxwell) give the same result. Furthermore, the potentials that occur in the effective action have a clear
  stringy interpretation. One can, then, express the charges in terms of the
  numbers of certain branes defining the background in which the string is
  quantized. These numbers occur in the resulting CFT whose states are counted
  and in the microscopic entropy. The comparison with the macroscopic one is,
  therefore, straightforward.

  When the $\alpha'$ corrections are switched on, though, the Chern-Simons
  terms in the field strengths become active and, on top of this,
  higher-derivative terms are added to the equations of motion of the
  potentials. The values of the conserved charges are then different for different
  definitions and their physical interpretation in terms of branes is far less
  clear. We believe this is a problem to which insufficient attention has been
  paid so far and which is crucial for having a convincing comparison of
  macroscopic and microscopic computations of black-hole entropies. We will
  investigate it in more depth elsewhere \cite{kn:ORZ} and, here, we will
  content ourselves with using simple and intuitive notions of charge.

\end{enumerate}

This paper is organized as follows: in Section~\ref{sec-heteroticalpha} we
review the Heterotic Superstring effective action (equations of motion and
supersymmetry transformations) to first order in $\alpha'$ in the
Bergshoeff-de Roo formulation \cite{Bergshoeff:1989de}.  In
Section~\ref{sec-d5} we focus on the 5-dimensional, static, 3-charge, extremal
black hole solutions of the theory, originally studied in Refs.~\cite{Callan:1996dv, Tseytlin:1996as, Cvetic:1996xz}. First of all, in Section~\ref{sec-d5-10dimensional}, we
present a 10-dimensional ansatz the encompasses all the solutions that we are
going to study.  In Section~\ref{sec-d5-susy} we study the unbroken
supersymmetries of the ansatz for different signs of the three charges and the
5-dimensional form of the ansatz obtained by compactification on a T$^{5}$ in
Section~\ref{sec-d5-5dimensional}. The $\alpha'$ corrections are presented in
Section~\ref{sec-d5-corrections} and the calculation of their Wald entropies
is carried out in Section~\ref{sec-d5-entropy}.

In Section~\ref{sec-d4} we focus on 4-dimensional, static, 4-charge, extremal
black hole solutions of the theory, which includes the heterotic version of
the black holes studied in
Refs.~\cite{Cvetic:1995kv, Maldacena:1996gb,Johnson:1996ga}. Again, we start by presenting in
Section~\ref{sec-d4-10dimensional} a general ansatz that encompasses all the
solutions that we are going to consider, in Section~\ref{sec-d4-susy} we study
its unbroken supersymmetry for different signs of the charges and in
Section~\ref{sec-d4-4dimensional} we give the 4-dimensional form of the
ansatz, obtained by compactification on a T$^{6}$. In
Section~\ref{sec-d4-corrections} we present the $\alpha'$ corrections, which
now depend very strongly on the relative signs of the charges. We write the
fields of the $\alpha'$-corrected solutions in a manifestly
T~duality-invariant form.  In Section~\ref{sec-d4-entropy} we present the
computation of the Wald entropy of the $\alpha'$-corrected solutions.

In Section~\ref{sec-Tduality} we explore the use of $\alpha'$-corrected
T~duality to help us determine the form of the $\alpha'$ corrections to the
solutions in the 5- and 4-dimensional cases considered here.  Finally, in
Section~\ref{sec-discussion} we discuss different aspects of the solutions
presented in this paper.

\section{The Heterotic Superstring effective action}
\label{sec-heteroticalpha}

For the sake of self-consistency, we are going to give a short description of
the Heterotic Superstring (HST) effective action and the fermionic
supersymmetry transformation rules to first order in $\alpha'$ as given in
\cite{Bergshoeff:1989de} using the conventions of
Ref.~\cite{Ortin:2015hya}.\footnote{The relation between the fields used here
  and those in Ref.~\cite{Bergshoeff:1989de} can be found in
  Ref.~\cite{Fontanella:2019avn}.} We remind the reader that
$\alpha'=\ell_{s}^{2}$, where $\ell_{s}$ is the string length.

The HST effective action describes the massless degrees of freedom of the HST:
the (string-frame) Zehnbein $e^{a}=e^{a}{}_{\mu}dx^{\mu}$, the Kalb-Ramond
2-form $B=\tfrac{1}{2}B_{\mu\nu}dx^{\mu}\wedge dx^{\nu}$, the dilaton $\phi$
and the Yang-Mills field $A^{A}=A^{A}{}_{\mu}dx^{\mu}$ (where $A,B,C,\ldots$ take
values in the Lie algebra of the gauge group). In certain cases we will use
hats to distinguish the 10-dimensional fields from those obtained from them by
dimensional reduction.

We start by defining the (torsionless, metric-compatible) Levi-Civita spin
connection 1-form $\omega^{a}{}_{b}=\omega_{\mu}{}^{a}{}_{b}dx^{\mu}$, which
in our conventions satisfies the Cartan structure equation in the form

\begin{equation}
\mathcal{D}e^{a}\equiv de^{a}-\omega^{a}{}_{b}\wedge e^{b}=0\,.
\end{equation}

The theory is, however, most conveniently described in terms of a pair of
torsionful spin (or Lorentz) connections
${\Omega}^{(0)}_{(\pm)}{}^{{a}}{}_{{b}}
={\Omega}^{(0)}_{(\pm)\,\mu}{}^{{a}}{}_{{b}}dx^{\mu}$ obtained by adding an
extra term to the Levi-Civita connection

\begin{equation}
  \label{eq:Omegapm0def}
{\Omega}^{(0)}_{(\pm)}{}^{{a}}{}_{{b}} 
=
{\omega}^{{a}}{}_{{b}}
\pm
\tfrac{1}{2}{H}^{(0)}_{{\mu}}{}^{{a}}{}_{{b}}dx^{{\mu}}\,.
\end{equation}

In this definition, $H^{(0)}_{{\mu}}{}^{{a}}{}_{{b}}$ are the components of
the zeroth order Kalb-Ramond 3-form field strength

\begin{equation}
H^{(0)} \equiv dB\,.
\end{equation}

The curvature 2-forms of these connections and the (Lorentz-) Chern-Simons
3-forms are defined as\footnote{Analogous formulae apply to the curvature
  2-form and Chern-Simons 3-form of the Levi-Civita connection.}

\begin{eqnarray}
{R}^{(0)}_{(\pm)}{}^{{a}}{}_{{b}}
& = & 
d {\Omega}^{(0)}_{(\pm)}{}^{{a}}{}_{{b}}
- {\Omega}^{(0)}_{(\pm)}{}^{{a}}{}_{{c}}
\wedge  
{\Omega}^{(0)}_{(\pm)}{}^{{c}}{}_{{b}}\,,
\\
& & \nonumber \\
{\omega}^{{\rm L}\, (0)}_{(\pm)}
& = &  
d{\Omega}^{ (0)}_{(\pm)}{}^{{a}}{}_{{b}} \wedge 
{\Omega}^{ (0)}_{(\pm)}{}^{{b}}{}_{{a}} 
-\tfrac{2}{3}
{\Omega}^{ (0)}_{(\pm)}{}^{{a}}{}_{{b}} \wedge 
{\Omega}^{ (0)}_{(\pm)}{}^{{b}}{}_{{c}} \wedge
{\Omega}^{ (0)}_{(\pm)}{}^{{c}}{}_{{a}}\,.  
\end{eqnarray}

The curvature 2-form and Chern-Simons 3-form of the Yang-Mills field are
defined, respectively, by

\begin{eqnarray}
{F}^{A}
& = & 
d{A}^{A}+\tfrac{1}{2} f_{BC}{}^{A}{A}^{B}\wedge{A}^{C}\,, 
\\
& & \nonumber \\
{\omega}^{\rm YM}
& = & 
dA_{A}\wedge {A}^{A}+\tfrac{1}{3}f_{ABC}{A}^{A}\wedge{A}^{B}\wedge{A}^{C}\,.
\end{eqnarray}

\noindent
In the last expression we have used the Killing metric of the gauge group's
Lie algebra in the relevant representation, $K_{AB}$ (which we assume to be
invertible and positive definite) to lower adjoint indices $A,B,C,\ldots$.

Now we can define the first-order Kalb-Ramond 3-form field strength as

\begin{equation}
\label{eq:H1def}
  H^{(1)}
= 
d{B}
+\frac{\alpha'}{4}\left({\omega}^{\rm YM}+{\omega}^{{\rm L}\, (0)}_{(-)}\right)\,.    
\end{equation}

\noindent
Replacing $H^{(0)}$ by $H^{(1)}$ in the definition of the torsionful spin
connections Eq.~(\ref{eq:Omegapm0def}) we obtain the first-order torsionful
spin connections ${\Omega}^{(1)}_{(\pm)}{}^{{a}}{}_{{b}}$ whose Chern-Simons
3-form ${\omega}^{{\rm L}\, (1)}_{(\pm)}$ can be used in Eq.~(\ref{eq:H1def})
to define the second-order Kalb-Ramond 3-form field strength $H^{(2)}$ etc. In
the first-order action we only need $H^{(1)}$,
${\Omega}^{(0)}_{(-)}{}^{{a}}{}_{{b}}$ and its curvature 2-form
$R^{(0)}_{(-)}{}^{a}{}_{b}$ and Chern-Simons 3-form
${\omega}^{{\rm L}\, (0)}_{(-)}$. The kinetic term for $H^{(1)}$ (quadratic)
will contain terms of second order in $\alpha'$ which contribute to the
equations of motion and which we should ignore. Keeping them, though, makes
the action explicitly gauge invariant instead of invariant up to terms of
second order in $\alpha'$.






The explicit $\alpha'$ corrections in the action, equations of motion and in
the Bianchi identity of the Kalb-Ramond 3-form field strength can be
conveniently described in terms of the so-called ``$T$-tensors'', defined by

\begin{equation}
\label{eq:Ttensors}
\begin{array}{rcl}
{T}^{(4)}
& \equiv &
\dfrac{\alpha'}{4}\left[
{F}_{A}\wedge{F}^{A}
+
{R}_{(-)}{}^{{a}}{}_{{b}}\wedge {R}_{(-)}{}^{{b}}{}_{{a}}
\right]\,,
\\
& & \\ 
{T}^{(2)}{}_{{\mu}{\nu}}
& \equiv &
\dfrac{\alpha'}{4}\left[
{F}_{A}{}_{{\mu}{\rho}}{F}^{A}{}_{{\nu}}{}^{{\rho}} 
+
{R}_{(-)\, {\mu}{\rho}}{}^{{a}}{}_{{b}}{R}_{(-)\, {\nu}}{}^{{\rho}\,  {b}}{}_{{a}}
\right]\,,
\\
& & \\    
{T}^{(0)}
& \equiv &
{T}^{(2)\,\mu}{}_{{\mu}}\,.
\\
\end{array}
\end{equation}

The  string-frame HST effective action is, to first order in $\alpha'$,

\begin{equation}
\label{heterotic}
S
=
\frac{g_{s}^{2}}{16\pi G_{N}^{(10)}}
\int d^{10}x\sqrt{|{g}|}\, 
e^{-2{\phi}}\, 
\left\{
{R} 
-4(\partial{\phi})^{2}
+\tfrac{1}{12}{H}^{(1)\, 2}
-\tfrac{1}{2}T^{(0)}
\right\}\,,
\end{equation}

\noindent
where $R$ is the Ricci scalar of the string-frame metric
$g_{\mu\nu}=\eta_{ab}e^{a}{}_{\mu}e^{b}{}_{\nu}$, $G_{N}^{(10)}$ is the
10-dimensional Newton constant , $g_{s}$ is the HST coupling constant (the
vacuum expectation value of the dilaton $e^{<\phi>}$ which we will identify
with the asymptotic value of the dilaton $e^{\phi_{\infty}}$ in
asymptotically-flat black-hole solutions).

The 10-dimensional Newton constant depends on the string length and coupling
constant in this way:

\begin{equation}
  \label{eq:10dNewtonconstant}
G_{N}^{(10)} = 8\pi^{6}g_{s}^{2}\ell_{s}^{8}\,.  
\end{equation}

The naive variation of the action Eq.~(\ref{heterotic}) leads to very
complicated equations of motion which contain terms with higher derivatives,
all of them coming from the variation of the torsionful spin connection in
different terms. However, the lemma proven in Ref.~\cite{Bergshoeff:1989de}
allows us to consistently ignore all those terms because they give
contributions of second and higher orders in $\alpha'$. Thus, the equations that
we effectively have to solve to first order in $\alpha'$ come from the
variation of the action with respect to the \textit{explicit}
(\textit{i.e.}~not in the torsionful spin connection) occurrences of the
fields. After some recombinations, they can be written in the standard form
found in the literature

\begin{eqnarray}
\label{eq:eq1}
  R^{\mu}{}_{a} -2\nabla^{\mu}\partial_{a}\phi
+\tfrac{1}{4}{H}^{\mu\rho\sigma}{H}_{a\rho\sigma}
  -T^{(2)\, \mu}{}_{a}
  & = &
0\,,
\\
& & \nonumber \\
\label{eq:eq2}
(\partial \phi)^{2} -\tfrac{1}{2}\nabla^{2}\phi
-\tfrac{1}{4\cdot 3!}{H}^{2}
+\tfrac{1}{8}T^{(0)}
  & = &
0\,,
\\
& & \nonumber \\
  \label{eq:eq3}
      \nabla_{\mu}\left(e^{-2\phi}H^{\mu\nu\rho}\right)
        & = &
0\,,
\\
& & \nonumber \\
  \label{eq:eq4}
\alpha' e^{2\phi}\nabla_{(+)\, \mu}\left(e^{-2\phi}F^{A\, \mu\nu}\right)
        & = &
0\,.
\end{eqnarray}

The Einstein equation can be rewritten in a more compact form using the
identity

\begin{equation}
  R^{\mu}{}_{a}(\Omega^{(\pm)})
  -2\nabla^{(\pm)\,\mu}\partial_{a}\phi
  =
  R^{\mu}{}_{a}(\omega)
  +\tfrac{1}{4}H^{\mu bc}H_{abc}
  -2\nabla^{\mu}\partial_{a}\phi
  \pm\tfrac{1}{2}e^{2\phi}\nabla^{\nu}\left(e^{-2\phi}H^{\mu\nu}{}_{a}\right)\,,
\end{equation}

\noindent
and combining it with the Kalb-Ramond equation:

\begin{equation}
  R^{\mu}{}_{a}(\Omega^{(\pm)})-2\nabla^{(\pm)\,\mu}\partial_{a}\phi
  -T^{(2)\, \mu}{}_{a}
  = 
0\,.  
\end{equation}

To first order in $\alpha'$ and for vanishing fermions, the supersymmetry
transformation rules of the gravitino $\psi_{\mu}$, dilatino $\lambda$ and
gaugini $\chi^{A}$ (all of them 32-component Majorana-Weyl spinors,
$\psi_{\mu},\chi^{A}$ and $\epsilon$ with positive chirality and $\lambda$ with negative
chirality) are

\begin{eqnarray}
  \label{eq:gravitino}
  \delta_{\epsilon} \psi_{a}
  & = &
        \nabla^{(+)}{}_{a}\,  \epsilon
        \equiv
        \left(\partial_{a}-\tfrac{1}{4} \Omega^{(+)}{}_{a \, bc} \Gamma^{bc}\right)
        \epsilon \,,
  \\
  \nonumber \\
  \label{eq:dilatino}
  \delta_{\epsilon} \lambda
  & = &
        \bigg( \partial_{a} \phi \Gamma^{a}
        -\tfrac{1}{12} H_{abc} \Gamma^{abc} \bigg) \epsilon \,,
  \\
  \nonumber \\
  \label{eq:gaugino}
  \alpha'\delta_{\epsilon} \chi^{A}
  & = &
        - \tfrac{1}{4} \alpha' F^{A}{}_{ab} \Gamma^{ab} \epsilon\,. 
\end{eqnarray}

\noindent
We will not include Yang-Mills fields in our solutions.

\section{3-charge, 5-dimensional black holes}
\label{sec-d5}

\subsection{10-dimensional form}
\label{sec-d5-10dimensional}
The extremal 3-charge, 5-dimensional black holes we want to study were originally obtained in \cite{Callan:1996dv, Tseytlin:1996as, Cvetic:1996xz}.
A common 10-dimensional ansatz for them is

\begin{subequations}
\label{eq:3-charge10dsolution}
\begin{align}
d\hat{s}^{2}
  & =
    \frac{1}{\mathcal{Z}_{+}\mathcal{Z}_{-}}dt^{2}
    -\mathcal{Z}_{0}(d\rho^{2}+\rho^{2}d\Omega_{(3)}^{2})
    \nonumber \\
  & \nonumber \\
  & 
    -\frac{k_{\infty}^{2}\mathcal{Z}_{+}}{\mathcal{Z}_{-}}
    \left[dz+\beta_{+}k_{\infty}^{-1}
    \left(\mathcal{Z}^{-1}_{+}-1\right)dt\right]^{2}
    -dy^{m}dy^{m}\,,
\hspace{.5cm}
m=1,\ldots,4\,,
\label{eq:d10metric}
  \\
& \nonumber \\
\hat{H}
& = 
 d\left[\beta_{-} k_{\infty}\left(\mathcal{Z}^{-1}_{-}-1\right)
 dt \wedge dz\right]
 +\beta_{0}\rho^{3}\mathcal{Z}'_{0}\omega_{(3)}\,,
\\
& \nonumber \\
e^{-2\hat{\phi}}
& = 
e^{-2\hat{\phi}_{\infty}}
\mathcal{Z}_{-}/\mathcal{Z}_{0}\,,
\end{align}
\end{subequations}

\noindent
where a prime indicates derivation with respect to the radial coordinate
$\rho$,

\begin{subequations}
  \begin{align}
  d\Omega^{2}_{(3)}
  & =
  \frac{1}{4}\left[ (d\psi+\cos{\theta}d\varphi)^{2}
    + d\Omega^{2}_{(2)} \right]\,,
    \\
    & \nonumber \\
  d\Omega^{2}_{(2)}
  & =
  d\theta^{2}+\sin^{2}{\theta}d\varphi^{2}\,,
  \end{align}
\end{subequations}

\noindent
are, respectively, the metrics of the round 3- and 2-spheres of unit radius,

\begin{equation}
  \label{eq:omega3}
\omega_{(3)}= \frac{1}{8}d\cos{\theta}\wedge d\varphi\wedge d\psi\,,
\end{equation}

\noindent
is the volume 3-form of the former and the functions
$\mathcal{Z}_{0},\mathcal{Z}_{+},\mathcal{Z}_{-}$ are functions of $\rho$ 
which have to be determined by imposing the equations of motion at the desired
order in $\alpha'$.

The zeroth-order equations of motion are solved when the functions
$\mathcal{Z}_{0},\mathcal{Z}_{+},\mathcal{Z}_{-}$ are given by 

\begin{equation}
\label{eq:Zs3-chargezerothorder}
\mathcal{Z}_{0,\pm}
= 
1+\frac{q_{0,\pm}}{\rho^{2}}\,.
\end{equation}

The three parameters $q_{0,\pm}$, which are related to the black-hole charges, are
assumed to be positive but, otherwise, arbitrary. The parameters
$\beta_{0\pm}$, which can take the values $\pm 1$, are related to the signs
of the charges. The solution also has two independent moduli: the asymptotic
value of the 10-dimensional dilaton, $\hat{\phi}_{\infty}$, and the radius of
the compact dimension parametrized by the coordinate $z$ measured in string-length units, $k_{\infty}\equiv R_{z}/\ell_{s}$.  The signs of the charges are
not indifferent: the supersymmetry of the solutions and the form of the
$\alpha'$ corrections depend very strongly on them, as we are going to
see.

\subsection{Supersymmetry}
\label{sec-d5-susy}

Les us determine the unbroken supersymmetries of the field configurations
described by the ansatz in Eqs.~(\ref{eq:3-charge10dsolution}) for arbitrary
choices of the functions
$\mathcal{Z}_{0},\mathcal{Z}_{+},\mathcal{Z}_{-}$.\footnote{See also
  Ref.~\cite{Chimento:2018kop}.}  We use the Zehnbein basis

\begin{equation}
  \begin{aligned}
    {\hat e}^{0} & = \frac{1}{\sqrt{\mathcal{Z}_{+}\mathcal{Z}_{-}}}dt\,,
    \hspace{.7cm}
    {\hat e}^{1} = \sqrt{\mathcal{Z}_{0}}d\rho\,,
    \hspace{.7cm}
    {\hat e}^{i+1} = \sqrt{\mathcal{Z}_{0}}\rho/2 v^{i}\,,
    \\
    & \\
    {\hat e}^{5}&  =
    k_{\infty}\sqrt{\frac{\mathcal{Z}_{+}}{\mathcal{Z}_{-}}}
    \left[dz+\beta_{+}k_{\infty}^{-1}
      \left(\mathcal{Z}^{-1}_{+}-1\right)dt\right]\,,
    \hspace{.7cm}
    {\hat e}^{m} = dy^{m}\,,
  \end{aligned}
\end{equation}

\noindent
where the $v^{i}$, $i=1,2,3$ are the SU$(2)$ left-invariant Maurer-Cartan
1-forms, satisfying the Maurer-Cartan equation

\begin{equation}
  \label{eq:SU2MC}
  dv^{i} = -\tfrac{1}{2}\epsilon^{ijk}v^{j}\wedge v^{k}\,.
  \hspace{1cm}
  \epsilon^{123}=+1\,,
\end{equation}

Plugging this configuration into the Killing spinor equations
$\delta_{\epsilon} \psi_{a}=0$, $\delta_{\epsilon} \lambda=0$ and
$\delta_{\epsilon} \chi^{A}=0$ with the supersymmetry variations in
Eqs.~(\ref{eq:gravitino})-(\ref{eq:gaugino}), we immediately see that the
third of them (the gaugini's) is automatically satisfied. It is not hard to
see that the second (the dilatino's) is satisfied for supersymmetry parameters
satisfying the two (compatible) conditions

\begin{subequations}
  \begin{align}
    \label{eq:susyprojector1}
    \tfrac{1}{2}\left(1-\beta_{-}\Gamma^{05} \right)\epsilon
    & =
    0\,,
    \\
    & \nonumber \\
    \label{eq:susyprojector2}
    \tfrac{1}{2}\left(1+\beta_{0}\Gamma^{1234} \right)\epsilon
    & =
    0\,.
  \end{align}
\end{subequations}

Solving $\delta_{\epsilon} \psi_{0}=0$ with a time-independent spinor, though,
demands $\beta_{-}=\beta_{+}$, and using this condition in
$\delta_{\epsilon} \psi_{1}=0$ we find that

\begin{equation}
\epsilon = (\mathcal{Z}_{+}\mathcal{Z}_{-})^{-1/4}\epsilon_{0}\,,  
\end{equation}

\noindent
where $\epsilon_{0}$ is an $r-$independent spinor that satisfies the above conditions.

The equations $\delta_{\epsilon} \psi_{i}=0$ with $i=2,3,4$ (the directions in
the 3-sphere) take the form

\begin{equation}
  \label{eq:KSEi1}
\left(v_{i}-\tfrac{1}{4}(1+\beta_{0})\Gamma^{i1}\right)\epsilon_{0}=0\,,
\end{equation}

\noindent
where $v_{i}$ are the vectors dual to the left-invariant Maurer-Cartan 1-forms
in SU$(2)$ (S$^{3}$), $v^{i}$ defined above in Eq.~(\ref{eq:SU2MC}).
When $\beta_{0}=-1$, $\epsilon_{0}$ is just a constant spinor. When
$\beta_{0}=+1$, we can rewrite the equation, as an equation on SU$(2)$, in
the form

\begin{equation}
  \label{eq:KSEi2}
\left(d-v^{i}T_{i}\right)\epsilon_{0}=0\,,
\end{equation}

\noindent
where we have defined the SU$(2)$ generators

\begin{equation}
  T_{i} \equiv \tfrac{1}{2}\Gamma^{i1}\,.
\end{equation}

\noindent
Indeed, it can be checked that they satisfy the commutation relations

\begin{equation}
  [T_{i},T_{j}] = -\varepsilon_{ijk}T_{k}\,,
\end{equation}

\noindent
in the subspace of spinors satisfying Eq.~(\ref{eq:susyprojector2}) with
$\beta_{0}=+1$. Since, by definition,\footnote{Here we are using the
  conventions and results of Ref.~\cite{AlonsoAlberca:2002gh}.}
$v^{i}T_{i}=-u^{-1}du$, where $u$ is a generic element of SU$(2)$,
Eq.~(\ref{eq:KSEi2}) is equivalent to

\begin{equation}
  \label{eq:KSEi3}
  d(u\epsilon_{0})=0\,,
  \,\,\,\,\,
  \Rightarrow
  \,\,\,\,\,
  \epsilon_{0} = u^{-1}\epsilon_{00}\,,
\end{equation}

\noindent
where $\epsilon_{00}$ may, at most, depend on $z$. However,
$\delta_{\epsilon} \psi_{5}=0$ is solved for $z$-independent $\epsilon_{00}$
upon use of the projector Eq.~(\ref{eq:susyprojector1}) with
$\beta_{-}=\beta_{+}$, and the rest of the Killing spinor equations,
$\delta_{\epsilon} \psi_{m}=0$, are trivially solved for $y^{m}$-independent
(\textit{i.e.}~constant) $\epsilon_{00}$.

The conclusion is that the field configurations with $\beta_{-}=\beta_{+}$
(and arbitrary $\beta_{0}=\pm 1$) are the only supersymmetric ones, although the Killing
spinors are quite different for $\beta_{0}=+1$ and $\beta_{0}=-1$ cases. This
result is true regardless of the values of the functions
$\mathcal{Z}_{0},\mathcal{Z}_{+},\mathcal{Z}_{-}$ which means that the
$\alpha'$ corrections that we are going to determine preserve the unbroken
supersymmetries of the zeroth-order solution.

\subsection{5-dimensional form}
\label{sec-d5-5dimensional}

Upon trivial dimensional reduction on T$^{4}$ (parametrized by the coordinates
$y^{m}$) we get a 6-dimensional field configuration which is identical to the
one described in Section~\ref{sec-d5-10dimensional} except for the absence of
the $-dy^{m}dy^{m}$ term in the metric. An additional non-trivial
compactification on the S$^{1}$ parametrized by the coordinate $z$ using the
relations in Appendix~\ref{app-dictionaryfirstorder} gives the 5-dimensional
field configuration\footnote{Observe that the definition of the 5-dimensional
  winding vector $B^{(1)}$, Eq.~(\ref{eq:B1}), includes explicit $\alpha'$
  corrections, independent of those of the $\mathcal{Z}$ functions, that we
  are going to compute.}${}^{,}$\footnote{The action that results from this
  dimensional reduction to leading order in $\alpha'$ is given in Eq.~(A.1) of
  Ref.~\cite{Elgood:2020mdx}. In particular, notice that, after
  compactification on a T$^{n}$, the $(10-n)$-dimensional string coupling
  constant $g_{s}^{(10-n)}$ and Newton constant $G_{N}^{(10-n)}$ are related
  to the 10-dimensional ones and to the volume of the compactification torus,
  $V_{n}$, by
  \begin{subequations}
    \label{eq:relationsbetweenconstants}
    \begin{align}
      g_{s}^{(10-n)} & = g_{s}/\sqrt{V_{n}/(2\pi\ell_{s})^{n}}\,,
      \\
                     & \nonumber \\
       G_{N}^{(10-n)} & = G_{N}^{(10)}/V_{n}\,.
    \end{align}
  \end{subequations}
  In this case, we are taking the radii of the coordinates $y^{m}$ that
  parametrize the circles of the T$^{4}$ to be 1 in $\ell_{s}$ units and the
  radius of the S$^{1}$ parametrized by $z$ to be $R_{z}$, so that
  $V_{5}= (2\pi\ell_{s})^{5}R_{z}/\ell_{s}= (2\pi\ell_{s})^{5}k_{\infty}$.
}

\begin{subequations}
\label{eq:5dsolution}
\begin{align}
ds^{2}
  & =
    \frac{1}{\mathcal{Z}_{+}\mathcal{Z}_{-}}dt^{2}
    -\mathcal{Z}_{0}(d\rho^{2}+\rho^{2}d\Omega_{(3)}^{2})\,,
\\
& \nonumber \\
H^{(1)}
& = 
\beta_{0}\rho^{3}\mathcal{Z}'_{0}\omega_{(3)}\,,
\\
& \nonumber \\
  A
  & =
   \beta_{+}k_{\infty}^{-1}
    \left(-1+\frac{1}{\mathcal{Z}_{+}}\right)dt \,,
  \\
& \nonumber \\
  B^{(1)}
  & =
    \beta_{-}k_{\infty}\left\{-1 +\frac{1}{\mathcal{Z}_{-}}
    \left[1
    +\frac{\alpha'}{4}(1+\beta_{+}\beta_{-})
    \frac{\mathcal{Z}_{+}'\mathcal{Z}_{-}'}{\mathcal{Z}_{0}\mathcal{Z}_{+}\mathcal{Z}_{-}}
    \right]\right\}dt\,,
  \\
& \nonumber \\
e^{-2\phi}
  & =
    e^{-2\phi_{\infty}}
\sqrt{\mathcal{Z}_{+}\mathcal{Z}_{-}}/\mathcal{Z}_{0}\,,
\\
& \nonumber \\
  k
  & =
   k_{\infty} \sqrt{\mathcal{Z}_{+}/\mathcal{Z}_{-}}\,.
\end{align}
\end{subequations}

The KR 2-form can be dualized into a 1-form $A^{0}$, whose field
strength, $F^{0}=dA^{0}$, is defined by

\begin{equation}
F^{0}=e^{-2\phi}\star H^{(1)}\,,  
\end{equation}

\noindent
and it is not difficult to see that, in a convenient gauge,

\begin{equation}
A^{0} = \beta_{0} e^{-2\phi_{\infty}}(\mathcal{Z}_{0}^{-1}-1)\,.
\end{equation}

\noindent
Finally, the scalar combination $k^{(1)}$ defined in Eq.~(\ref{eq:k1def}) is given by

\begin{equation}
  \label{eq:k15d}
  k^{(1)}
  =
  k_{\infty}\sqrt{\mathcal{Z}_{+}/\mathcal{Z}_{-}}
  \left\{
    1+\frac{\alpha'}{4}(1+\beta_{+}\beta_{-})
    \frac{\mathcal{Z}_{+}'\mathcal{Z}_{-}'}{\mathcal{Z}_{0}\mathcal{Z}_{+}\mathcal{Z}_{-}}
    \right\}\,.
\end{equation}

\noindent
The (modified) Einstein-frame metric \cite{Maldacena:1996ky} is

\begin{equation}
  \label{eq:5dmetric}
ds_{E}^{2}
  =
f^{2}dt^{2}
-f^{-1}(d\rho^{2}+\rho^{2}d\Omega_{(3)}^{2})\,,
\hspace{1.5cm}
f^{-3}= \mathcal{Z}_{+}\mathcal{Z}_{-}\mathcal{Z}_{0}\,.
\end{equation}

When the functions $\mathcal{Z}_{0,\pm}$ are given by
Eqs.~(\ref{eq:Zs3-chargezerothorder}) and we remove the explicit $\alpha'$
correction in the winding vector $B^{(1)}$, this field configuration is a
solution of the zeroth-order in $\alpha'$ equations of motion that describes
an asymptotically-flat, static, extremal black hole whose event horizon lies
at $\rho=0$. In this limit the metric takes the characteristic
AdS$_{2}\times$S$^{3}$ form. The radii of the two factors is equal to
$(q_{+}q_{-}q_{0})^{1/6}$ and the Bekenstein-Hawking (BH) entropy is

\begin{equation}
  \label{eq:BHentropy3charge1}
  S
  =
  \frac{A}{4G_{N}^{(5)}}
  =
  \frac{\pi^{2}}{2G_{N}^{(5)}}\sqrt{q_{+}q_{-}q_{0}}\,,  
\end{equation}

\noindent
From the asymptotic expansion of the metric at spatial infinity
($\rho\rightarrow \infty$) we can read the mass

\begin{equation}
  M
  =
  \frac{\pi}{4G_{N}^{(5)}}\left(q_{+}+q_{-}+q_{0}\right)\,.
\end{equation}

\noindent
The 5-dimensional charges can be defined as

\begin{subequations}\label{eq:5dchargesdef}
  \begin{align}
    \mathcal{Q}_{+}
    & =
      \frac{g_{s}^{(5)\,2}\ell_{s}}{16\pi G_{N}^{(5)}}
      \int_{S^{3}_{\infty}} e^{-2\phi}k^{2}\star F\,,
    \\
    & \nonumber \\
    \mathcal{Q}_{-}
    & =
      \frac{g_{s}^{(5)\,2}\ell_{s}}{16\pi G_{N}^{(5)}}
      \int_{S^{3}_{\infty}} e^{-2\phi}k^{-2}\star G^{(0)}\,,
    \\
    & \nonumber \\
    \mathcal{Q}_{0}
    & =
      \frac{g_{s}^{(5)\,2}\ell_{s}}{16\pi G_{N}^{(5)}}
      \int_{S^{3}_{\infty}} H^{(0)}\,,
  \end{align}
\end{subequations}

\noindent
where the factors of $\ell_{s}$ have been introduced so that they are dimensionless.
Their values are found to be

\begin{equation}
  \label{eq:charges}
\mathcal{Q}_{+}
=
\frac{R_{z}^{2}\beta_{+}q_{+}}{g_{s}^{2}\ell_{s}^{4}}\,,
\hspace{1cm}
\mathcal{Q}_{-}
=
\frac{\beta_{-}q_{-}}{g_{s}^{2}\ell_{s}^{2}}\, ,\\
\hspace{1cm}
\mathcal{Q}_{0}
=
-\frac{\beta_{0}q_{0}}{\ell_{s}^{2}}\,,
\end{equation}

\noindent
after use of Eqs.~(\ref{eq:10dNewtonconstant}) and
(\ref{eq:relationsbetweenconstants}). In fact, defined in this way, these charges are quantized and represent
the number of different stringy objects:  $\mathcal{Q}_{+}$ represents the momentum of a wave along the compact direction $z$, $\mathcal{Q}_{-}$ is winding number of a fundamental string wrapping that direction, and $\mathcal{Q}_{0}$ stands for the
number of solitonic (NS) 5-branes.  These three dimensionless numbers can be
positive or negative depending on the constants $\beta_{0,\pm}$. In
terms of these numbers, the mass and BH entropy are given by

\begin{subequations}
  \begin{align}
    M
    & =
    \frac{1}{R_{z}}|\mathcal{Q}_{+}|+\frac{R_{z}}{\ell_{s}^{2}}|\mathcal{Q}_{-}|+\frac{R_{z}}{\ell_{s}^{2}g_{s}^{2}}|\mathcal{Q}_{0}|\,,
    \\
    & \nonumber \\
    S
    & =
      2\pi\sqrt{|\mathcal{Q}_{+}\mathcal{Q}_{-}\mathcal{Q}_{0}|}\, .
  \end{align}
\end{subequations}

\subsection{$\alpha'$ corrections}
\label{sec-d5-corrections}

As shown in
Refs.~\cite{Cano:2018qev,Chimento:2018kop,Cano:2018brq,Cano:2019ycn}, it is
much easier to find the $\alpha'$-corrected solution working directly in 10
dimensions since the 10-dimensional action contains less fields and takes a simpler form.
We compute the $T$-tensors using the
zeroth-order solution and make an ansatz for the corrections based on the form
of the $T$-tensors. In the present case, as suggested by our previous results,
the ansatz has exactly the same form as the original solution in
Eqs.~(\ref{eq:3-charge10dsolution}) but now we assume that the $\mathcal{Z}$
functions can get corrections which we have to determine by solving the
equations. More explicitly, we assume that the $\mathcal{Z}$ functions are
given by

\begin{equation}
  \mathcal{Z}_{\pm,0}
  =
  1 +\frac{q_{\pm,0}}{\rho^{2}} +\alpha'\delta
  \mathcal{Z}_{\pm,0}(\rho)\,,  
\end{equation}

\noindent
and we just have to find the functions $\delta \mathcal{Z}_{\pm,0}(\rho)$.

Observe that the same functions occur in different fields and they might be
corrected in different forms when they are part of different fields. In other words, the ansatz in terms if the $\mathcal{Z}_{\pm,0}$ functions
might not be general enough for the corrected solutions, as it contains too few independent functions. In fact, there is no reason to expect that the form of the ansatz will be preserved except for supersymmetric solutions, since supersymmetry can strongly constrain the form of a given solution, see for instance \cite{Papadopoulos:2008rx, Ruiperez:2020qda, Cano:2021dyy}.
However, for all our five-dimensional solutions, the ansatz (\ref{eq:3-charge10dsolution}) turns out to solve all the equations. 
This is not the case for the 4-dimensional solutions that we
are going to consider later, which do need a more general ansatz.

The differential equations for the corrections
$\delta \mathcal{Z}_{\pm,0}(\rho)$ have been solved demanding, as boundary
conditions, that the value at spatial infinity of the original function (1 in
all cases) is not modified and, furthermore, that the coefficient of the
$1/\rho^{2}$ term in the $\rho\rightarrow 0$ (near-horizon) expansion is not
modified, either. The result is

\begin{subequations}
\label{eq:Zs3-chargefirstorder}
\begin{align}
\delta\mathcal{Z}_{0}
& = 
-\frac{\rho^{2}+2q_{0}}{(\rho^{2}+q_{0})^{2}}\,,
\\
& \nonumber \\
\delta\mathcal{Z}_{+}
& = 
(1+\beta_{+}\beta_{-})
\frac{q_{+}(\rho^{2}+q_{0}+q_{-})}{q_{0}(\rho^{2}+q_{-})(\rho^{2}+q_{0})}\,,
\\
& \nonumber \\
\delta\mathcal{Z}_{-}
& =
0\,.
\end{align}
\end{subequations}

\noindent
It is also interesting to rewrite the corrected $\mathcal{Z}$s as follows:

\begin{subequations}
\label{eq:Zs3-chargefirstorder-rewritten}
\begin{align}
  \label{eq:Z0correctedd5}
  \mathcal{Z}_{0}
  & = 
    1+\frac{\hat{q}_{0}}{\rho^{2}}
    +\alpha' \frac{\hat{q}_{0}^{2}}{\rho^{2}(\rho^{2}+\hat{q}_{0})^{2}}
    +\mathcal{O}(\alpha^{\prime\,2})
  \\
  & \nonumber \\
  \label{eq:Z+correctedd5}
  \mathcal{Z}_{+}
  & = 
    1+\frac{\hat{q}_{+}}{\rho^{2}}
    -\alpha'(1+\beta_{+}\beta_{-})\frac{\hat{q}_{+}\hat{q}_{-}}{\rho^{2}(\rho^{2}+\hat{q}_{-})(\rho^{2}+\hat{q}_{0})}
    +\mathcal{O}(\alpha^{\prime\,2})\,,
  \\
  & \nonumber \\
  \mathcal{Z}_{-}
  & =
    1+\frac{\hat{q}_{-}}{\rho^{2}}+\mathcal{O}(\alpha^{\prime\,2})\,,
\end{align}
\end{subequations}

\noindent
where we have defined, for convenience,

\begin{subequations}
  \label{eq:asymptoticcoefficients}
  \begin{align}
   \hat{q}_{0}
    & \equiv
    q_{0}-\alpha'\,,
    \\
    & \nonumber \\
    \hat{q}_{+}
    & \equiv
      q_{+}\left[1+\alpha'\frac{(1+\beta_{+}\beta_{-})}{q_{0}}\right]\,,
    \\
    & \nonumber \\
    \hat{q}_{-}
    & \equiv
      q_{-}\,,
  \end{align}
\end{subequations}

\noindent
because they are the coefficients of the $1/\rho^{2}$ terms in the asymptotic
expansion. Then, not surprisingly, the mass is now given by

\begin{equation}
  \begin{aligned}
    M
    &=
    \frac{\pi}{4G_{N}^{(5)}}\left(\hat{q}_{+}+\hat{q}_{-}+\hat{q}_{0}\right)
    \\
    & \\
    & =
    \frac{\pi}{4G_{N}^{(5)}}\left\{q_{+}
      +q_{-}
      +q_{0}
      +\alpha'\left[1+(1+\beta_{+}\beta_{-})\frac{q_{+}}{q_{0}}\right]\right\}\,.
  \end{aligned}
\end{equation}

\noindent
However, $A/(4G_{N}^{(5)})$ is still given by Eq.~(\ref{eq:BHentropy3charge1})
because of the boundary conditions we have chosen. This is not the entropy of
the corrected black hole anymore, though, as we will see in the next section.

It is interesting to compute the explicit form of the corrected Kaluza-Klein
and winding vector fields $A_{\mu}$ and $B^{(1)}{}_{\mu}$:

\begin{subequations}
  \begin{align}
    A_{\mu}
    & =
      \beta_{+}\delta^{t}{}_{\mu}k^{-1}_{\infty}
  \left\{-1+\frac{\rho^{2}}{\rho^{2}+\hat{q}_{+}}\left[1
      +\alpha'
      (1+\beta_{+}\beta_{-})\frac{\hat{q}_{+}\hat{q}_{-}}{(\rho^{2}+\hat{q}_{+})(\rho^{2}+\hat{q}_{-})(\rho^{2}+\hat{q}_{0})}\right]\right\}\,,
    \\
    & \nonumber \\
    \label{eq:B1}
    B^{(1)}{}_{\mu}
  & =
  \delta^{t}{}_{\mu}\beta_{-}k_{\infty}
  \left\{-1+\frac{\rho^{2}}{\rho^{2}+\hat{q}_{-}}\left[1
      +\alpha'
      (1+\beta_{+}\beta_{-})\frac{\hat{q}_{+}\hat{q}_{-}}{(\rho^{2}+\hat{q}_{+})(\rho^{2}+\hat{q}_{-})(\rho^{2}+\hat{q}_{0})}\right]\right\}
\,,
  \end{align}
\end{subequations}

\noindent
as well as the corrected Kaluza-Klein scalars $k$ and $k^{(1)}$ (defined in
Eq.~(\ref{eq:k1def})):

\begin{subequations}
  \begin{align}
    k
    & =
      k_{\infty} \sqrt{\frac{\rho^{2}+\hat{q}_{+}}{\rho^{2}+\hat{q}_{-}}}
      \left[1
      -\alpha'
      \frac{(1+\beta_{+}\beta_{-})}{2}\frac{\hat{q}_{+}\hat{q}_{-}}{(\rho^{2}+\hat{q}_{+})(\rho^{2}+\hat{q}_{-})(\rho^{2}+\hat{q}_{0})}\right]\,,
    \\
    & \nonumber \\
    k^{(1)}
    & =
      k_{\infty} \sqrt{\frac{\rho^{2}+\hat{q}_{+}}{\rho^{2}+\hat{q}_{-}}}
      \left[1
      +\alpha'
      \frac{(1+\beta_{+}\beta_{-})}{2}\frac{\hat{q}_{+}\hat{q}_{-}}{(\rho^{2}+\hat{q}_{+})(\rho^{2}+\hat{q}_{-})(\rho^{2}+\hat{q}_{0})}\right]\,.
  \end{align}
\end{subequations}

As discussed in \cite{Elgood:2020xwu}, the $\alpha'$-corrected T~duality rules
derived in \cite{Bergshoeff:1995cg} are equivalent to the transformations

\begin{equation}
  \label{eq:1storderTduality}
  A_{\mu}'
  =
  B^{(1)}{}_{\mu}\,, 
  \hspace{1cm}
  B^{(1)}{}_{\mu}'
  = 
  A_{\mu}\,,
  \hspace{1cm}
  k'
   = 
  1/k^{(1)}\,,
\end{equation}

\noindent
of the compactified theory (here, along the S$^{1}$ parametrized by $z$). The
above expressions show that the effect of the T~duality transformation on this
solution is equivalent to the following transformation of the parameters of
the solution:

\begin{equation}
  \label{eq:Tdualitytransformationoftheparametersd=5}
  \beta_{\pm}'= \beta_{\mp}\,,
  \hspace{1cm}
  \hat{q}_{\pm}'= \hat{q}_{\mp}\,,
  \hspace{1cm}
  k_{\infty}'=1/k_{\infty}\,.
\end{equation}

\noindent
In terms of these variables it is clear that the mass of the
$\alpha'$-corrected black holes is invariant under T~duality to first order in
$\alpha'$.

In terms of the original parameters $q_{\pm}$, the T~duality
transformations read

\begin{equation}
  \beta_{\pm}'= \beta_{\mp}\,,
  \hspace{1cm}
  q_{\pm}'= q_{\mp}
  \left(1\mp\alpha'\frac{(1+\beta_{+}\beta_{-})}{q_{0}}\right)\,,
  \hspace{.7cm}
  k_{\infty}'=1/k_{\infty}\,,
\end{equation}

\noindent
and the invariance of the mass is not so obvious, although it still holds to
first order in $\alpha'$.

The relation between the charge parameters $q_{0,\pm}$ or $\hat{q}_{0,\pm}$
and the conserved charges of the system is much more complicated than in the
zeroth-order case, due to the non-linearity of the action and the amount of
terms in which a given field occurs. The definitions of the conserved charges
and their physical meaning will be investigated elsewhere \cite{kn:ORZ}. 
For the purposes of this paper, we find it useful to work with the following set of charges, defined as in the zeroth-order case (but replacing the field strengths by their first-order expressions),

\begin{subequations}\label{eq:5dchargesdeffirstorder}
  \begin{align}
    \hat{\mathcal{Q}}_{+}
    & =
      \frac{g_{s}^{(5)\,2}\ell_{s}}{16\pi G_{N}^{(5)}}
      \int_{S^{3}_{\infty}} e^{-2\phi}k^{2}\star F=
\beta_{+}\frac{k^2_{\infty}{\hat q}_{+}}{g_s^2 \ell_{s}}\,,
    \\
    & \nonumber \\
    \hat{\mathcal{Q}}_{-}
    & =
      \frac{g_{s}^{(5)\,2}\ell_{s}}{16\pi G_{N}^{(5)}}
      \int_{S^{3}_{\infty}} e^{-2\phi}k^{-2}\star G^{(1)}=\beta_{-}\frac{\hat{q}_{-}}{g_{s}^{2}\ell_{s}^{2}}\,,
    \\
    & \nonumber \\
    \hat{\mathcal{Q}}_{0}
    & =
      \frac{g_{s}^{(5)\,2}\ell_{s}}{16\pi G_{N}^{(5)}}
      \int_{S^{3}_{\infty}} H^{(1)}=-\beta_{0}\frac{\hat{q}_{0}}{\ell_{s}^{2}}\,,
  \end{align}
\end{subequations}
We call them  $\hat{\mathcal{Q}}_{0\pm}$ to distinguish them from the zeroth-order charges $\mathcal{Q}_{0\pm}$, and to emphasize that, with this definition, they are proportional to the hatted parameters $\hat q_{0\pm}$ that control the asymptotic behaviour of the $\mathcal{Z}_{0\pm}$ functions (and hence, of the gauge fields). 

\subsection{$\alpha'$-corrected entropy}
\label{sec-d5-entropy}

The gauge-invariant Wald-entropy formula for heterotic black-hole solutions has been
recently derived in Ref.~\cite{Elgood:2020nls}. In 10-dimensional language, it
can be written as an integral over a spatial section of the
horizon\footnote{Strictly speaking, this formula has been derived assuming the
  black hole is non-extremal and, originally, it takes the form of an integral
  over the bifurcation sphere. Using the arguments of
  Ref.~\cite{Jacobson:1993vj}, we can extrapolate this formula to the extremal
  case, expressing it as an integral over any spatial section of the horizon.}
that we denote by $\Sigma$

\begin{equation}
  \label{eq:Waldentropyformula}
  S
  =
   \frac{g_{s}^{2}}{8G_{N}^{(10)}}
 \int_{\Sigma}
 e^{-2\hat\phi}
 \left\{
   \left[
    \hat \star ({\hat e}^{\hat a}\wedge {\hat e}^{\hat b})
        +\frac{\alpha'}{2}\star \hat R^{(0)}_{(-)}{}^{\hat a \hat b}
    \right]{\hat n}_{\hat a\hat b}
          +(-1)^{d}\frac{\alpha'}{2}\Pi_{n}\wedge \hat\star {\hat H}^{(0)}
        \right\}\,,
\end{equation}

\noindent
where the 1-form $\Pi_{n}$ is the \textit{vertical Lorentz momentum map}
associated to the binormal to the Killing horizon, $\hat n^{\hat a\hat b}$

\begin{equation}
  d\Pi_{n}
 \stackrel{\Sigma}{=}
  {\hat R}^{(0)}_{(-)}{}^{\hat a\hat b}{\hat n}_{\hat a\hat b}\,.
\end{equation}

In all the solutions that we are considering (supersymmetric or not), the
following property is satisfied \cite{Prester:2008iu}:

\begin{equation}
  {\hat R}^{(0)}_{(-)}{}^{\hat a\hat b}  
  \stackrel{\cal H}{=}
  0\,.
\end{equation}

\noindent
(Observe that the spatial section of the horizon $\Sigma$ is
S$^{3}\times$T$^{5}$.) Furthermore, in the frame that we are using, \cite{Elgood:2020nls}

\begin{equation}
  \Pi_{n}
  \stackrel{\cal H}{=}
  {\hat \Omega}^{(0)}_{(-)}{}^{\hat a\hat b}{\hat n}_{\hat a\hat b}\,.
\end{equation}

\noindent
so that, for the 10-dimensional solutions under consideration and in the particular
frame that we are using, the entropy is effectively given by the integral

\begin{equation}
  \label{eq:Waldentropyformula2}
  S
  =
   \frac{{g}_{s}^{2}}{8G_{N}^{(10)}}
 \int_{\Sigma}
 e^{-2\hat \phi}
 \left\{
     {\hat\star} ({\hat e}^{\hat a}\wedge {\hat e}^{\hat b}){\hat n}_{\hat a\hat b}
        -\frac{\alpha'}{2}\hat\star {\hat H}^{(0)} \wedge {\hat\Omega}^{(0)}_{(-)}{}^{\hat a\hat b}{\hat n}_{\hat a\hat b}
        \right\}\,.
\end{equation}

It is worth stressing that this integral is not Lorentz-invariant but we only
claim it to be valid for the kind of Lorentz frames we are using. On the other
hand, had we used the Iyer-Wald prescription \cite{Iyer:1994ys}, we would have
obtained an almost identical formula, only differing from this one by the
value of the coefficient of the second term: $-\alpha'/4$ instead of
$-\alpha'/2$. However, the formula is, as a matter of fact, radically
different from Eq.~(\ref{eq:Waldentropyformula2}), because in the Iyer-Wald
case it is supposed to be valid in \textit{any} reference frame and it is
clearly not.

Let us proceed to evaluate this integral on the 10-dimensional solution that
corresponds to the 5-dimensional $\alpha'$-corrected black holes.

The first term in this integral gives the Bekenstein-Hawking contribution
(\textit{i.e.},~the area of the horizon measured with the (modified) Einstein
frame metric):\footnote{The normalization of the binormal is such that
  $ {\hat n}^{\hat a\hat b}{\hat n}_{\hat a\hat b}=-2$. Also, $ \omega_{(3)}$ is the S$^{3}$ volume form in
  Eq.~(\ref{eq:omega3}).}

\begin{equation}
  \begin{aligned}
    \frac{{g}_{s}^{2}}{8G_{N}^{(10)}} \int_{\Sigma} e^{-2\hat \phi} \hat\star
    ({\hat e}^{\hat a}\wedge {\hat e}^{\hat b}){\hat n}_{\hat a\hat b}
    & =
    \frac{1}{4G_{N}^{(10)}}\int_{\Sigma}d^{8}S e^{-2(\hat \phi-{\hat \phi}_{\infty)}}
    \\
    & \\
    & =
    \frac{1}{4G_{N}^{(10)}}\lim_{\rho\rightarrow 0}
    \int \omega_{(3)} dz dy^{1}\cdots dy^{4}
    k_{\infty}\sqrt{\rho^{6}\mathcal{Z}_{+}\mathcal{Z}_{-}\mathcal{Z}_{0}}
    \\
    & \\
    & =
    \frac{1}{4G_{N}^{(10)}} 2\pi^{2} (2\pi\ell_{s})^{5}k_{\infty}
    \sqrt{q_{+}q_{-}q_{0}}
    \\
    & \\
    & =
    \frac{\pi^{2}}{2G_{N}^{(5)}}\sqrt{q_{+}q_{-}q_{0}}\,,
  \end{aligned}
\end{equation}

\noindent
which looks identical to the result we obtained at zeroth order,
Eq.~(\ref{eq:BHentropy3charge1}). It is, however, different, if we express it
in terms of the coefficients of the $1/\rho^{2}$ terms in the asymptotic
expansions, defined in Eqs.~(\ref{eq:asymptoticcoefficients}).

As for the second term, since the only non-vanishing component of the
binormal in our conventions is ${\hat n}^{01}=+1$, we have

\begin{equation}
  \begin{aligned}
    -\frac{{g}_{s}^{2}\alpha'}{16G_{N}^{(10)}}\int_{\Sigma}
    e^{-2\hat \phi}\hat \star {\hat H}^{(0)} \wedge {\hat\Omega}^{(0)}_{(-)}{}^{\hat a\hat b}{\hat n}_{\hat a\hat b}
    & = 
    \frac{\alpha'}{8G_{N}^{(10)}} \int_{\Sigma}
    e^{-2(\hat \phi-{\hat\phi}_{\infty})}\hat\star {\hat H}^{(0)} \wedge {\hat\Omega}^{(0)}_{(-)}{}^{01}
    \\
    & \\
    & =
    \frac{\alpha'}{8G_{N}^{(10)}} \int_{\Sigma}
    e^{-2(\hat \phi-{\hat \phi}_{\infty})}\hat\star {\hat H}^{(0)} \wedge
    \left({\hat\Omega}^{(0)}_{(-)\,0}{}^{01}{\hat e}^{0}
     +{\hat \Omega}^{(0)}_{(-)\,5}{}^{01}{\hat e}^{5}\right)\,.
   \end{aligned}
\end{equation}

Since the integral is over a spatial section, the first term does not
contribute to the integral and we have
 
\begin{equation}
  \begin{aligned}
    &\hspace{.5cm}
    -\frac{\alpha'}{8G_{N}^{(10)}} \int_{\Sigma}
     \omega_{(3)}\wedge dy^{1}\wedge
    \cdots dy^{4} \wedge dz
    \left(\mathcal{Z}_{0}^{1/2}\rho\right)^{3}
    k_{\infty}\left(\mathcal{Z}_{+}/\mathcal{Z}_{-}\right)^{1/2}
    e^{-2(\phi-\phi_{\infty})}{\hat H}^{(0)}_{015}{\hat \Omega}^{(0)}_{(-)\,501}
    \\
    & \\
    & =
    \frac{\alpha' \pi^{2} (1+\beta_{+}\beta_{-})}{2G_{N}^{(5)}}
    \sqrt{\frac{q_{+}q_{-}}{q_{0}}}\,,
   \end{aligned}
\end{equation}

\noindent
so

\begin{equation}
S =
\frac{\pi^{2}}{2G_{N}^{(5)}}\sqrt{q_{+}q_{-}q_{0}}
\left[1+\alpha'\frac{(1+\beta_{+}\beta_{-})}{q_{0}}\right]\,.
\end{equation}
\noindent
Our result agrees with previous literature in the supersymmetric ($\beta_{+}\beta_{-}=+1$) \cite{Prester:2008iu, Faedo:2019xii,Elgood:2020xwu, Cano:2021dyy} and non-supersymmetric ($\beta_{+}\beta_{-}=-1$) cases \cite{Prester:2008iu}. Furthermore, in the supersymmetric case, it agrees with the microscopic calculation performed in Ref.~\cite{ Castro:2008ys}. This agreement is more evident when one expresses the entropy in terms of the charges of the solution, which is what we do now. To this aim, we first rewrite it in terms of ${\hat q}_{+}$ and ${\hat q}_0$ as follows,

\begin{equation}
  \begin{aligned}
    S & =
    \frac{\pi^{2}}{2G_{N}^{(5)}}\sqrt{q_{+}\left[1+\alpha'\frac{(1+\beta_{+}\beta_{-})}{q_{0}}\right]q_{-}q_{0}\left[1+\alpha'\frac{(1+\beta_{+}\beta_{-})}{q_{0}}\right]}
    \\
    & \\
    & = \frac{\pi^{2}}{2G_{N}^{(5)}}\sqrt{\hat{q}_{+}q_{-}
      \left[\hat{q}_{0}+\alpha'(2+\beta_{+}\beta_{-})\right]}\,.
  \end{aligned}
\end{equation}

\noindent
Finally, by using the charges introduced in Eq.~(\ref{eq:5dchargesdeffirstorder}), we can recast the entropy in the following form:

\begin{equation}
S
=
2\pi
\sqrt{|\hat{\mathcal{Q}}_{+}\hat{\mathcal{Q}}_{-}|
\left(|\hat{\mathcal{Q}}_{0}|+2+\beta_{+}\beta_{-}\right)}\,.
\end{equation}

\section{4-charge, 4-dimensional black holes}
\label{sec-d4}

\subsection{10-dimensional form}
\label{sec-d4-10dimensional}

The extremal 4-charge 4-dimensional black holes that we want study in this section were first considered in \cite{Cvetic:1995kv, Maldacena:1996gb, Johnson:1996ga}. A common 10-dimensional ansatz to describe them is

\begin{subequations}
\label{eq:4-charge10dsolution}
  \begin{align}
d\hat{s}^{2}
    & =
          \frac{1}{\mathcal{Z}_{+}\mathcal{Z}_{-}}dt^{2}
-\mathcal{Z}_{0}\mathcal{H}\left(dr^{2}+r^{2}d\Omega^{2}_{(2)}\right)
      \nonumber \\
    & \nonumber \\
    & \hspace{.5cm}
      -\ell_{\infty}^{2}\frac{\mathcal{Z}_{0}}{\mathcal{H}}
      \left(dw+\beta\ell_{\infty}^{-1}q\cos{\theta} d\varphi\right)^{2}
    \nonumber \\
  & \nonumber \\
  & \hspace{.5cm}
    -k_{\infty}^{2}\frac{\mathcal{Z}_{+}}{\mathcal{Z}_{-}}
    \left[dz +\beta_{+}k_{\infty}^{-1}
    \left(\mathcal{Z}^{-1}_{+}-1\right)dt\right]^{2}
      -dy^{i}dy^{i}\,,
\\
&  \nonumber \\
\hat{H}
    & =
       d\left[ \beta_{-}k_{\infty}\left(\mathcal{Z}^{-1}_{-}-1\right)
 dt \wedge dz\right]
      +\beta_{0}\ell_{\infty}r^{2}\mathcal{Z}_{0h}'\,
      \omega_{(2)}\wedge dw \,,  
\\
& \nonumber \\
    e^{2\hat\phi}
    & =
-\frac{C_{\phi}\mathcal{Z}_{0}\mathcal{Z}_{-}'r^{2}}{\mathcal{Z}_{-}q_{-}}\, ,
  \end{align}
\end{subequations}

\noindent
where a prime indicates a derivative with respect to the radial coordinate
$r$, $C_{\phi}$ is a constant to be adjusted so that the asymptotic value of
the dilaton is $e^{2\hat{\phi}_{\infty}}$,\footnote{The ansatz for the dilaton
  has a form convenient to handle the $\alpha'$ corrections because it solves
  automatically the Kalb-Ramond equation. It reduces to the standard
  $C_{\phi}\mathcal{Z}_{0}/\mathcal{Z}_{-}$ with $C_{\phi}=e^{2\phi_{\infty}}$
  at zeroth order in $\alpha'$ and whenever $\mathcal{Z}_{-}$ gets no
  $\alpha'$ corrections.}

\begin{equation}
  \omega_{(2)}
  =
\sin{\theta}  d\theta\wedge d\varphi\,,
\end{equation}

\noindent
is the volume form of the round 2-sphere of unit radius, and the 5 functions
$\mathcal{Z}_{0}$, $\mathcal{Z}_{0h}$, $\mathcal{Z}_{+}$, $\mathcal{Z}_{-}$
and $\mathcal{H}$ depend on the radial coordinate $r$ only. These
functions are determined by imposing the equations of motion to the desired
order in $\alpha'$ and, to zeroth order, they must take the form

\begin{subequations}
\label{eq:Zs4-charge}
\begin{align}
\mathcal{Z}_{0,\pm}
& = 
1+\frac{q_{0,\pm}}{r}\,, 
\\
& \nonumber \\
\mathcal{Z}_{0h}
& = 
\mathcal{Z}_{0}\,,
\\
& \nonumber \\
\mathcal{H}
& = 
1+\frac{q}{r}
\end{align}
\end{subequations}

\noindent
Again, the four constants $q_{0,\pm},q$, which are associated to the four independent
charges of the 4-dimensional solution, are assumed to be positive. The
constants $\beta_{0,\pm},\beta$ take care of the signs of the associated
charges.  The solution depends on three independent moduli: the asymptotic value
of the 10-dimensional dilaton $\hat{\phi}_{\infty}$ and the radii of the
compact dimensions parametrized by the coordinates $z$ and $w$ measured in
string-length units $k_{\infty}\equiv R_{z}/\ell_{s}$ and
$\ell_{\infty} \equiv R_{w}/\ell_{s}$.

\subsection{Supersymmetry}
\label{sec-d4-susy}

It is important to determine which of the field configurations described by
the general ansatz Eq.~(\ref{eq:4-charge10dsolution}) are supersymmetric. We
use the Zehnbein basis

\begin{equation}\label{eq:basis4d}
  \begin{aligned}
    {\hat e}^{0} & = \frac{1}{\sqrt{\mathcal{Z}_{+}\mathcal{Z}_{-}}}dt\,,
    \hspace{.7cm}
    {\hat e}^{1} = \sqrt{\mathcal{Z}_{0}\mathcal{H}}\, dr\,,
    \hspace{.7cm}
    {\hat e}^{2} = \sqrt{\mathcal{Z}_{0}\mathcal{H}}\, r v^{2}\,,
    \hspace{.7cm}
    {\hat e}^{3} = \sqrt{\mathcal{Z}_{0}\mathcal{H}}\, r v^{1}\,,
    \\
    & \\
    {\hat e}^{4}&  =
    \ell_{\infty}\sqrt{\frac{\mathcal{Z}_{0}}{\mathcal{H}}}
    \left[dw+\beta \ell_{\infty}^{-1}q\cos{\theta}d\varphi \right]\,,
    \\
    & \\
    {\hat e}^{5}&  =
    k_{\infty}\sqrt{\frac{\mathcal{Z}_{+}}{\mathcal{Z}_{-}}}
    \left[dz+\beta_{+}k_{\infty}^{-1}
      \left(\mathcal{Z}^{-1}_{+}-1\right)dt\right]\,,
    \hspace{.7cm}
    {\hat e}^{m} = dy^{m}\,,
  \end{aligned}
\end{equation}

\noindent
where

\begin{equation}
  v^{1} =\sin{\theta} d\varphi\,,
  \hspace{1cm}
  v^{2} = d\theta\,,
\end{equation}

\noindent
are horizontal components of the left-invariant Maurer-Cartan 1-form of the
SU$(2)/$U$(1)$ coset space.\footnote{The details of this construction are
  given in Appendix~\ref{app:coset}.}

The dilatino Killing spinor equation (KSE) is $\delta_{\epsilon} \lambda = 0$
where the supersymmetry variation of the dilatino with vanishing fermions is
given in Eq.~(\ref{eq:dilatino}). Substituting the values of the fields, it
can be brought to the form

\begin{equation}
  \left\{
    \frac{\mathcal{Z}_{0}'}{\mathcal{Z}_{0}}
    \left[1+\beta_{0}\frac{\mathcal{Z}_{0h}'}{\mathcal{Z}_{0}'}\Gamma^{1234}\right]
    -\frac{\mathcal{Z}_{-}'}{\mathcal{Z}_{-}}
    \left[1-\beta_{-}\Gamma^{05}\right]
    +\left(\frac{\mathcal{Z}_{-}''}{\mathcal{Z}_{-}'}+\frac{2}{r}\right)
  \right\}\epsilon =0\,.
\end{equation}

We can solve this equation without demanding any relations between
$\mathcal{Z}_{0}$ and $\mathcal{Z}_{-}$ if we demand the following two
conditions on the functions:

\begin{subequations}
  \begin{align}
    \mathcal{Z}_{0h}'
    & =
      \mathcal{Z}_{0}'\,,
    \\
    & \nonumber \\
    \mathcal{Z}_{-}'
    & \propto
      1/r^{2}\,,
  \end{align}
\end{subequations}

\noindent
which are satisfied by the zeroth-order solutions, and the following
conditions on the Killing spinors:

\begin{subequations}
  \begin{align}
    \label{eq:4d5braneprojector}
    \left[1+\beta_{0}\Gamma^{1234}\right]\epsilon
    & =
      0\,,
    \\
    & \nonumber \\
    \label{eq:4dwaveprojector}
    \left[1-\beta_{-}\Gamma^{05}\right] \epsilon
    & =
      0\,,
  \end{align}
\end{subequations}

\noindent
which are compatible and reduce the number of independent components of the
spinors to a $1/4$ of the total 16 real components.

The zeroth component of the gravitino KSE $\delta_{\epsilon} \psi_{a}=0$,
where the supersymmetry variation of the gravitino with vanishing fermions
is given in Eq.~(\ref{eq:gravitino}), can be brought to the form 

\begin{equation}
\left\{\partial_{0}
  -\frac{\left[\log{(\mathcal{Z}_{+}\mathcal{Z}_{-})}\right]'}{4(\mathcal{Z}_{0}\mathcal{H})^{1/2}}\left[1-\beta_{-}\frac{\left[\log{(\mathcal{Z}_{+}^{\beta_{+}\beta_{-}}\mathcal{Z}_{-})}\right]'}{\left[\log{(\mathcal{Z}_{+}\mathcal{Z}_{-})}\right]'}\Gamma^{05}\right]\right\}\epsilon
=
0\,,
\end{equation}

\noindent
and can be solved by a spinor satisfying

\begin{equation}
\partial_{0}\epsilon=0\,,  
\end{equation}

\noindent
and the condition (\ref{eq:4dwaveprojector}) if

\begin{equation}
  \label{eq:b+b-=1}
\beta_{+}\beta_{-}=+1\,.  
\end{equation}

The $a=1$ component of the gravitino KSE takes the form

\begin{equation}
  \left\{
    \partial_{r}
    +\tfrac{1}{4}\beta_{-}
    \left[\log{(\mathcal{Z}_{+}^{\beta_{+}\beta_{-}}\mathcal{Z}_{-})}\right]' \Gamma^{05}
  \right\}\epsilon
  =
  0\,,
\end{equation}

\noindent
and, after use of the conditions (\ref{eq:4dwaveprojector}) and (\ref{eq:b+b-=1}) it can be
solved by

\begin{equation}
\epsilon = (\mathcal{Z}_{+}\mathcal{Z}_{-})^{-1/4}\epsilon_{0}\,,  
\end{equation}

\noindent
where $\epsilon_{0}$ is an $r$-independent spinor that satisfies the conditions
Eqs.~(\ref{eq:4d5braneprojector}) and (\ref{eq:4dwaveprojector}).

The $a=4$ component takes the form

\begin{equation}
  \left\{
    \partial_{4}
    -\tfrac{1}{2}\Gamma^{14}\Omega_{(+)\, 414}
    \left[1+\beta_{0}
      \frac{\left[\log{(\mathcal{Z}_{0}/\mathcal{H})}\right]'}
      {\left[\frac{\mathcal{Z}_{0h}'}{\mathcal{Z}_{0}}+\beta\beta_{0}\frac{q/r^{2}}{\mathcal{H}}\right]}\Gamma^{1234}\right]
  \right\}\epsilon
  =
  0\,,
\end{equation}

\noindent
and, if we demand

\begin{subequations}
  \begin{align}
    \beta\beta_{0}
    & =
      +1\,,
    \\
    & \nonumber \\
    \mathcal{H}'
    & =
      -q/r^{2}\,,  
  \end{align}
\end{subequations}

\noindent
it is solved by a spinor satisfying Eq.~(\ref{eq:4d5braneprojector}) and

\begin{equation}
\partial_{4}\epsilon=0\,.  
\end{equation}

The $a=5$ component can be written in the form

\begin{equation}
  \left\{
    \partial_{5}
    -\tfrac{1}{2}\Gamma^{01}\Omega_{(+)\, 501}
    \left[1-\beta_{-}
      \frac{\left[\log{(\mathcal{Z}_{+}/\mathcal{Z}_{-})}\right]'}
      {\left[\log{(\mathcal{Z}_{+}^{\beta_{+}\beta_{-}}/\mathcal{Z}_{-})}\right]'}
      \Gamma^{05}
      \right]
  \right\}\epsilon
  =
  0\,,
\end{equation}

\noindent
and, if the conditions (\ref{eq:b+b-=1}) and
(\ref{eq:4dwaveprojector}) are satisfied, then it is solved if, in addition,

\begin{equation}
\partial_{5}\epsilon=0\,.  
\end{equation}

Finally, using all the conditions derived so far, the $a=2,3$ equations can be
combined into a single differential equation for the spinor
$\epsilon_{0}=\epsilon_{0}(\theta,\phi)$\footnote{The last 4 components of the
  gravitino KSE are trivially solved by $y$-independent spinors and the
  conditions $\partial_{0,4,5}\epsilon=0$ imply that $\epsilon_{0}$ is
  independent of the coordinates $r$, $t$, $w$, $z$ and $y^{i}$.}

\begin{equation}
  \label{eq:angularKSE}
  \left\{d +\tfrac{1}{2}\Gamma^{13}\sin{\theta} d\varphi
    +\tfrac{1}{2}\Gamma^{12}d\theta
    +\tfrac{1}{2}\Gamma^{23}\cos{\theta} d\varphi\right\}\epsilon_{0}
  =
  0\,.
\end{equation}

\noindent
We can make the following identifications with the generators of the su$(2)$
algebra

\begin{equation}
  P_{1} = \tfrac{1}{2}\Gamma^{13}\,,
  \hspace{1cm}
  P_{2} = \tfrac{1}{2}\Gamma^{12}\,,
  \hspace{1cm}
  M = \tfrac{1}{2}\Gamma^{23}\,,  
\end{equation}

\noindent
because they satisfy the commutation relations Eq.~(\ref{eq:su2algebrasplit}).
Then, using the components of the Maurer-Cartan 1-form in
Eq.~(\ref{eq:componentsMC1-form}), the KSE (\ref{eq:angularKSE}) can be
rewritten in the form

\begin{equation}
  \label{eq:angularKSE2}
  \left\{d +P_{a}v^{a}+M\vartheta \right\}\epsilon_{0}
  =
  0\,.
\end{equation}

\noindent
The 1-forms in this equation can be seen to add up to the left-invariant
Maurer-Cartan 1-form $V=-u^{-1}du$ (Eq.~(\ref{eq:MC1-form})) and the equation
can be rewritten in the form

\begin{equation}
  \label{eq:angularKSE3}
  \left\{d -u^{-1}du\right\}\epsilon_{0}
  =
  -u^{-1}d(u\epsilon_{0})=0\,.
\end{equation}

\noindent
Thus, it is solved by

\begin{equation}
  \epsilon_{0}
  =
  u^{-1}\epsilon_{00}
  =
  e^{(\theta-\pi/2)\tfrac{1}{2}\Gamma^{12}}   e^{\varphi\tfrac{1}{2}\Gamma^{13}}\epsilon_{00}\,,
\end{equation}

\noindent
where $\epsilon_{00}$ is a constant spinor that satisfies the conditions
(\ref{eq:4d5braneprojector}) and (\ref{eq:4dwaveprojector}).

Taking into account that we are considering configurations that, at zeroth
order in $\alpha'$ are determined by the functions given in
Eqs.~(\ref{eq:Zs4-charge}) which may have additional corrections at the next
order, we can summarize our results as follows: only the configurations of the
form Eq.~(\ref{eq:4-charge10dsolution}) which satisfy all the conditions

\begin{equation}
  \beta_{+}=\beta_{-}\,,
  \hspace{.5cm}
  \beta_{0}=\beta\,,
  \hspace{.5cm}
  \mathcal{Z}_{0h} = \mathcal{Z}_{0}\,,
  \hspace{.5cm}
  \mathcal{Z}_{-} = 1+\frac{q_{-}}{r}\,,
    \hspace{.5cm}
\mathcal{H} = 1+\frac{q}{r}\,,
\end{equation}

\noindent
are supersymmetric and their Killing spinors take the form

\begin{equation}
  \epsilon
  =
  (\mathcal{Z}_{+}\mathcal{Z}_{-})^{-1/4}
  e^{(\theta-\pi/2)\tfrac{1}{2}\Gamma^{12}}
  e^{\varphi\tfrac{1}{2}\Gamma^{13}}
  \epsilon_{00}\,,  
\end{equation}

\noindent
where the constant spinor $\epsilon_{00}$ satisfies

\begin{equation}
    \left[1+\beta_{0}\Gamma^{1234}\right]\epsilon_{00} = 0\,,
\hspace{1cm}
    \left[1-\beta_{-}\Gamma^{05}\right] \epsilon_{00} = 0 \,.
\end{equation}

\subsection{4-dimensional form}
\label{sec-d4-4dimensional}

Upon trivial dimensional reduction on T$^{4}$ (parametrized by the coordinates
$y^{i}$) of the ansatz in Eq.~(\ref{eq:4-charge10dsolution}) we get a
6-dimensional field configuration which is identical except for the absence of
the $-dy^{i}dy^{i}$ term in the metric. Further non-trivial compactification
on the T$^{2}$ parametrized by the coordinates $z$ and $w$ gives\footnote{This
  dimensional reduction can be carried out by using the results of
  Ref.~\cite{Ortin:2020xdm} or performing two consecutive dimensional
  reductions on circles using the results of Ref.~\cite{Elgood:2020xwu},
  reproduced in Appendix~\ref{app-dictionaryfirstorder}. In this particular case,
  the latter procedure is simpler.}${}^{,}$\footnote{In order to determine
  $B^{(1)}{}_{m\, \mu}$ we first have to integrate the KR field strength to
  find the components of the 10-dimensional KR 2-form potential
  $\hat{B}_{\hat{\mu}\hat{\nu}}$ and, in particular, the component
  $\hat{B}_{\varphi w}$.  In order to determine the integration constant in
  $\hat{B}_{\varphi w}$, we have assumed that the constant part of
  $r^{2}\mathcal{Z}_{0h}'$ at infinity is $-(q_{0}+\alpha'\delta q_{0})$. $\delta q_{0}$
  will be determined when we solve the equations of motion.}

\begin{subequations}
  \label{eq:4dimensionalansatz}
  \begin{align}
 ds^{2}
    & =
\frac{1}{\mathcal{Z}_{+}\mathcal{Z}_{-}}dt^{2}
-\mathcal{Z}_{0}\mathcal{H}\left(dr^{2}+r^{2}d\Omega^{2}_{(2)}\right)\,,
    \\
   & \nonumber \\
    H^{(1)}
    & =
      0\,,
    \\
    & \nonumber \\
   \left(A^{m}{}_{\mu}dx^{\mu}\right)
   & =
   \left(
     \begin{array}{c}
       \beta q/\ell_{\infty}\cos{\theta}d\varphi \\
       \\
       \beta_{+}k^{-1}_{\infty}\left(\mathcal{Z}^{-1}_{+}-1\right)dt \\
     \end{array}
    \right)\,,
    \\
  &  \nonumber \\
  \left(B^{(1)}{}_{m\, \mu}dx^{\mu}\right)
    & =
      \nonumber \\
    & \nonumber \\
&    \hspace{-3cm}
    \left(
    \beta_{0}\ell_{\infty}(q_{0}+\alpha'\delta q_{0h})
\cos{\theta}d\varphi\,,
    \,\,\,\,
    \beta_{-}k_{\infty}\left\{-1 +\frac{1}{\mathcal{Z}_{-}}
    \left[1+\frac{\alpha'}{4}(1+\beta_{+}\beta_{-})\frac{\mathcal{Z}_{+}'\mathcal{Z}_{-}'}{\mathcal{Z}_{+}\mathcal{Z}_{-}\mathcal{Z}_{0}\mathcal{H}}\right]\right\}  dt
    \right)\,,
    \\
    & \nonumber \\
      \left(G_{mn}\right)
  & =
  \left(
    \begin{array}{cc}
     \ell_{\infty}^{2}\mathcal{Z}_{0}/\mathcal{H}  & 0 \\
                                                            & \\
      0 & k_{\infty}^{2}\mathcal{Z}_{+}/\mathcal{Z}_{-} \\
     \end{array}
    \right)\,,
    \hspace{1cm}
    \left(B^{(1)}{}_{mn}\right)
    =
    0\,,
    \\
    & \nonumber \\
      e^{2\phi}
    & =
      -C_{\phi}k^{-1}_{\infty}\ell_{\infty}^{-1}\frac{\mathcal{Z}_{-}'r^{2}}{q_{-}}
  \sqrt{\frac{\mathcal{Z}_{0}\mathcal{H}}{\mathcal{Z}_{+}\mathcal{Z}_{-}}}\,.
  \end{align}
\end{subequations}

At zeroth order in $\alpha'$ or whenever $\mathcal{Z}_{-}$ gets no corrections, setting
$C_{\infty}=e^{2\hat{\phi}_{\infty}}$ and

\begin{equation}
e^{-2\phi_{\infty}} =  e^{-2\hat{\phi}_{\infty}}k_{\infty}\ell_{\infty}\,,
\end{equation}

\noindent
the ansatz for the dilaton takes the standard form

\begin{equation}
  e^{2\phi}
  =
  e^{2\phi_{\infty}}\sqrt{\frac{\mathcal{Z}_{0}\mathcal{H}}{\mathcal{Z}_{+}\mathcal{Z}_{-}}}\,.
\end{equation}

\noindent
The two scalar combinations $k^{(1)}$ and $\ell^{(1)}$ associated to the two
internal directions $z$ and $w$ and defined by Eq.~(\ref{eq:k1def}) applied to
these two directions are given by

\begin{subequations}
  \begin{align}
    k^{(1)}
    & =
      k_{\infty}\sqrt{\frac{\mathcal{Z}_{+}}{\mathcal{Z}_{-}}}
      \left[
      1+\frac{\alpha'}{4}(\beta_{+}\beta_{-}+1)\frac{\mathcal{Z}_{+}'\mathcal{Z}_{-}'}{
      \mathcal{Z}_{+}\mathcal{Z}_{-}\mathcal{Z}_{0}\mathcal{H}}
      \right]\,,
    \\
    & \nonumber \\
    \ell^{(1)}
    & =
      \ell_{\infty}\sqrt{\frac{\mathcal{Z}_{0}}{\mathcal{H}}}
      \left\{1-\frac{\alpha'}{4}\frac{1}{(\mathcal{Z}_{0}\mathcal{H})^{3}}
      \left[(\mathcal{Z}_{0}'\mathcal{H})^{2}+(\mathcal{Z}_{0}\mathcal{H}')^{2}
      +(\beta\beta_{0}-1)\mathcal{Z}_{0}'\mathcal{H}'\mathcal{Z}_{0}\mathcal{H} \right]
      \right\}
      \,.
  \end{align}
\end{subequations}

\noindent
The (modified) Einstein-frame metric is given by

\begin{equation}
  \label{eq:d4modifiedEinsteinframemetric}
  \begin{aligned}
    ds_{E}^{2}
    & =
    F\left[e^{-2U}dt^{2} -e^{2U}\left(dr^{2}+r^{2}d\Omega^{2}_{(2)}\right)\right]\,,
    \\
    & \\
    e^{2U} & =
    \sqrt{\mathcal{Z}_{+}\mathcal{Z}_{-}\mathcal{Z}_{0}\mathcal{H}}\,,
    \\
    & \\
    F^{-1} & =
    -C_{\phi}k^{-1}_{\infty}\ell_{\infty}^{-1}e^{-2\phi_{\infty}}\frac{\mathcal{Z}_{-}'r^{2}}{q_{-}}
    \,,
  \end{aligned}
\end{equation}

\noindent
and, in spite of its appearance, by construction it is asymptotically flat
with the standard normalization at spatial infinity if the string-frame metric
is. Observe that, for a non-trivial function $F(r)$, there is no redefinition
of the radial coordinate that makes this metric fit into the general extremal
black hole ansatz used in the FGK formalism \cite{Ferrara:1997tw} which is the
basis for most of the research on extremal black-hole attractors.\footnote{As
  it is well known, a generic 4-dimensional static spherically-symmetric
  metric is characterized by two independent functions of the radial
  coordinate. The ansatz used in Ref.~\cite{Ferrara:1997tw} for extremal and
  non-extremal black holes contains only one arbitrary function of $r$. This
  ansatz certainly encompasses most of the known asymptotically-flat
  extremal black-hole solutions, but there is no clear reason why it should encompass
  them all, as it has been implicitly assumed in most of the literature on
  black-hole attractors.}

At zeroth order in $\alpha'$ the equations of motion are solved by the
functions $\mathcal{Z}_{\pm,0},\mathcal{H}$ given in
Eqs.~(\ref{eq:Zs4-charge}) and, assuming all the constants $q_{\pm,0},q$ are
positive, the metric describes a regular extremal black hole whose event
horizon lies at $r=0$.\footnote{When some of those parameters are not
  positive, as a general rule, the solution has naked singularities at leading
  order in $\alpha'$ which, sometimes, can be removed by the first-order
  corrections as in Ref.~\cite{Cano:2018aod}, where a regular black hole which
  works as a one-way wormhole connecting two asymptotically-flat universes
  arises.} Close to the horizon the metric takes the characteristic
AdS$_{2}\times$S$^{2}$ form, where the metrics of both factors have radii
equal to $(q_{+}q_{-}q_{0}q)^{1/4}$. The Bekenstein-Hawking entropy is given
by

\begin{equation}
S= \frac{A}{4G_{N}^{(5)}} = \frac{\pi}{G_{N}^{(4)}}\sqrt{q_{+}q_{-}q_{0}\,q}\,,  
\end{equation}

\noindent
and the mass is given by

\begin{equation}
M = \frac{1}{4G_{N}^{(4)}}\left(q_{+}+q_{-}+q_{0}+q\right)\,.
\end{equation}

\noindent
The physically relevant charges of the system are in this case defined as

\begin{subequations}
  \label{eq:charges4dcase0}
\begin{align}
\mathcal{Q}_{+}
  & \equiv
  \frac{g_{s}^{(4)}\ell_{s}}{16\pi G_{N}^{(4)}} \int_{S^{2}_{\infty}}
  e^{-2\phi}k^{2}\star F^{z}
  =
  \beta _{+} \frac{2   \ell_{\infty} k_{\infty}^{2}}{\ell_{s}g_{s}^{2}} q_{+}\,,
  \\
  &\nonumber \\
\mathcal{Q}_{-}
  & \equiv 
  \frac{g_{s}^{(4)}\ell_{s}}{16\pi G_{N}^{(4)}} \int_{S^{2}_{\infty}}
  e^{-2\phi}k^{-2}\star G^{(0)}{}_{z}
  =
 \beta _{-} \frac{2  \ell_{\infty}}{\ell_{s} g_{s}^{2}}q_{-}\,,
  \\
  &\nonumber \\
\mathcal{Q}_{0}
  & \equiv
  \frac{g_{s}^{(4)}\ell_{s}}{16\pi G_{N}^{(4)}} \int_{S^{2}_{\infty}} G^{(0)}{}_{w}
  =
   \beta_{0}\frac{2 \ell_{\infty}}{\ell_{s}}q_{0}\,,
  \\
  & \nonumber\\
\mathcal{Q}
  & \equiv
  \frac{g_{s}^{(4)}\ell_{s}}{16\pi G_{N}^{(4)}} \int_{S^{2}_{\infty}} F^{w}
  =
  \beta \frac{2}{\ell_{\infty}\ell_{s}} q\,,
\end{align}
\end{subequations}
where the factor of $\ell_{s}$ ensures that these are dimensionless. At zeroth
order in $\alpha'$, these charges are quantized and they represent the winding number
$(\mathcal{Q}_{-})$ of a fundamental string, the momentum of the wave $(\mathcal{Q}_{+}$,
the number of S5-branes $(\mathcal{Q}_{0}$ and the topological charge
of the KK monopole $(\mathcal{Q})$. In terms of these, the zeroth-order mass and entropy read

\begin{subequations}
  \begin{align}
  M
  & =
  \frac{k_{\infty}}{\ell_s}|\mathcal{Q}_{-}|+\frac{1}{k_{\infty}\ell_{s}}|\mathcal{Q}_{+}|
  +\frac{k_{\infty}}{g_{s}^{2}\ell_{s}}|\mathcal{Q}_{0}|
    +\frac{\ell_{\infty}^{2}k_{\infty}}{g_{s}^{2}\ell_{s}}|\mathcal{Q}|\,,
    \\
    & \nonumber \\
    S
    & =
      2\pi 
      \sqrt{|\mathcal{Q}_{+}\mathcal{Q}_{-}\mathcal{Q}_{0}\mathcal{Q}|}\,.
  \end{align}
 \end{subequations}

\subsection{$\alpha'$ corrections}
\label{sec-d4-corrections}

The $\alpha'$ corrections to these solutions depend very strongly on the
choice of signs of the charges, \textit{i.e.}~on the choice of constants
$\beta_{\pm,0},\beta$. In particular, they are very different for the
$\beta=\beta_{0}$ and $\beta=-\beta_{0}$ cases, which have to be studied
separately. We remind the reader that the first case is supersymmetric if
$\beta_{+}=\beta_{-}$ while the second is always non-supersymmetric.

\subsubsection{Case 1:  \texorpdfstring{$\beta=\beta_{0}$}{beta equal
    beta0 } }
\label{sec-d4-corrections-extremal-alpha=alpha0andqneqq0}

In this case in which the S5-brane and KK monopole charges have the same sign the corrections take a relatively simple form:

\begin{subequations}
\label{eq:deltaZs4-chargecase1.2}
\begin{align}
\delta \mathcal{Z}_{+}
  & =
\frac{(\beta_{+}\beta_{-}+1)q_{+}}{4qq_{0}(r+q_{0})(r+q_{-})(r+q)}
    \left[r^{2}+(q_{0}+q+q_{-})r +qq_{0}+q_{-}q_{0} +qq_{-}\right]\,, 
\\
& \nonumber \\
\delta \mathcal{Z}_{-}
& = 
0\,, 
\\
& \nonumber \\
\delta \mathcal{Z}_{0,0h}
  & =
    -\frac{q^{2}+(2 q+r) (q+r)}{4 q (q+r)^{3}}
    -\frac{q_{0}^{2}+(2 q_{0}+r) (q+r)}{4 q (q_{0}+r)^{2} (q+r)}\,, 
\\
& \nonumber \\
\delta \mathcal{H}
& = 
0\,, 
  & \nonumber \\
  \label{Cphi1}
  C_{\phi}
 & =
  e^{2\hat{\phi}_{\infty}}\,.
\end{align}
\end{subequations}

\noindent
Note that, since in this case there are no corrections to $\mathcal{Z}_{-}$,
the expression of the dilaton can be simplified to

\begin{equation}
  e^{2\hat\phi}
  =
  e^{2\hat{\phi}_{\infty}}\frac{\mathcal{Z}_{0}}{\mathcal{Z}_{-}}\,,
\end{equation}

\noindent
and the function $F(r)$ that occurs in the modified Einstein frame metric
Eq.~(\ref{eq:d4modifiedEinsteinframemetric}) equals $1$. 

We observe that, in the supersymmetric case, corresponding to
$\beta_{+}\beta_{-}=1$, we recover the solution of
Ref.~\cite{Cano:2018brq}. On the other hand, just like for the 5-dimensional
black holes, the correction to $\mathcal{Z}_{+}$ vanishes identically for the
non-supersymmetric black holes $\beta_{+}\beta_{-}=-1$.

As we did in the 5-dimensional case, we can define the hatted parameters
$\hat{q}_{\pm,0,0h},\hat{q}$ as the coefficients of $1/r$ in the asymptotic
expansions of the corrected functions $\mathcal{Z}_{\pm,0,0h},\mathcal{H}$.\footnote{Note that not all these parameters can be identified with charges. In the case of $q_0$, the relevant charge parameter is $\hat{q}_{0h}$, which is the one appearing in the asymptotic expansion of $\mathcal{Z}_{0h}$. This is the function that controls the winding vector $B^{(1)}{}_{w\, \mu}$, rather than $\mathcal{Z}_{0}$. On the other hand, the asymptotic coefficient of the $1/r$ term in the function $\mathcal{H}$, which we call $\hat{q}$, is not a charge. Instead, the corresponding charge is related to the coefficient that appears in the KK vector $A^{w}{}_{\mu}$, which is always $q$.} In
this case, since $\mathcal{Z}_{0}=\mathcal{Z}_{0h}$, we have
$\hat{q}_{0}=\hat{q}_{0h}$ and we will denote them both by $\hat{q}_{0}$.  In
terms of the unhatted charge parameters, they are given by

\begin{subequations}
  \begin{align}
    \hat{q}_{+}
    & =
      q_{+}\left[1+\alpha'\frac{(\beta_{+}\beta_{-}+1)}{4qq_{0}}\right]\,,
    \\
    &  \nonumber \\
    \hat{q}_{-}
    & =
      q_{-}\,,
    \\
    & \nonumber \\
    \hat{q}_{0}
    & =
\hat{q}_{0h}= q_{0}-\frac{\alpha'}{2q}\,,
    \\
    & \nonumber \\
    \hat{q}
    & =
      q\,.
  \end{align}
\end{subequations}

\noindent
This determines $\delta q_{0}$ in $B^{(1)}{}_{w\, \varphi}$ to be

\begin{equation}
\delta q_{0}= -\frac{1}{2q}\,.  
\end{equation}

In terms of the hatted charge parameters, the corrected functions
$\mathcal{Z}_{\pm,0},\mathcal{H}$ take the form

\begin{subequations}
  \begin{align}
    \mathcal{Z}_{+}
    & =
      1+\frac{\hat{q}_{+}}{r}
      -\frac{\alpha'}{4}(\beta_{+}\beta_{-}+1)
      \frac{\hat{q}_{+}\hat{q}_{-}}{r(r+\hat{q}_{0})(r+\hat{q})(r+\hat{q}_{-})}\,,
    \\
    & \nonumber \\
    \mathcal{Z}_{-}
    & =
      1+\frac{\hat{q}_{-}}{r}\,,
    \\
    & \nonumber \\
    \mathcal{Z}_{0}
    & =
{\cal Z}_{0h}= 1+\frac{\hat{q}_{0}}{r}+\frac{\alpha'}{4}\frac{\hat{q}^{\,2}(r+\hat{q}_{0})^{2}+\hat{q}_{0}^{2}(r+\hat{q})^{2}}{r(r+\hat{q}_{0})^{2}(r+\hat{q})^{3}}\,,
    \\
    & \nonumber \\
    \mathcal{H}
    & =
      1+\frac{\hat{q}}{r}\,.
  \end{align}
\end{subequations}

Again, it is interesting to write explicitly the corrected Kaluza-Klein and
winding vectors which should be interchanged by T~duality in the $w$ and $z$
directions, respectively

\begin{subequations}
  \begin{align}
    A^{w}{}_{\mu}
    & =
      \beta \delta^{\varphi}{}_{\mu} q/\ell_{\infty}\cos{\theta}\,,
    \\
    & \nonumber \\
    B^{(1)}{}_{w\, \mu}
    & =
      \beta_{0} \delta^{\varphi}{}_{\mu}\ell_{\infty}\hat{q}_{0}\cos{\theta}\,,
  \end{align}
\end{subequations}
and

\begin{subequations}
  \begin{align}
    A^{z}{}_{\mu}
    & =
      \beta_{+}k_{\infty}^{-1}\left\{-1+\frac{r}{r+\hat{q}_{+}}\left[1
      +\frac{\alpha'}{4}(\beta_{+}\beta_{-}+1)
      \frac{\hat{q}_{+}\hat{q}_{-}}{(r+\hat{q}_{+})(r+\hat{q}_{-})(r+\hat{q}_{0})(r+\hat{q})}\right]\right\}\,,
    \\
    & \nonumber \\
    B^{(1)}{}_{z\, \mu}
    & =
      \beta_{-}k_{\infty}\left\{-1+\frac{r}{r+\hat{q}_{-}}\left[1
      +\frac{\alpha'}{4}(\beta_{+}\beta_{-}+1)
      \frac{\hat{q}_{+}\hat{q}_{-}}{(r+\hat{q}_{+})(r+\hat{q}_{-})(r+\hat{q}_{0})(r+\hat{q})}\right]\right\}\,,
  \end{align}
\end{subequations}

\noindent
as well as the scalars which are interchanged under the same transformations,
respectively

\begin{subequations}
  \begin{align}
    \ell
    & =
      \ell_{\infty}
      \sqrt{\frac{(r+\hat{q}_{0})}{(r+\hat{q})}}
      \left\{1
      +\frac{\alpha'}{8}\frac{\hat{q}^{\,2}(r+\hat{q}_{0})^{2}+\hat{q}_{0}^{2}(r+\hat{q})^{2}}{(r+\hat{q}_{0})^{3}(r+\hat{q})^{3}}\right\}\,,
    \\
    & \nonumber \\
    \ell^{(1)}
    & =
      \ell_{\infty}
      \sqrt{\frac{(r+\hat{q}_{0})}{(r+\hat{q})}}
      \left\{1
      -\frac{\alpha'}{8}\frac{\hat{q}^{\,2}(r+\hat{q}_{0})^{2}+\hat{q}_{0}^{2}(r+\hat{q})^{2}}{(r+\hat{q}_{0})^{3}(r+\hat{q})^{3}}\right\}\,,
  \end{align}
\end{subequations}

\noindent
and

\begin{subequations}
  \begin{align}
    k
    & =
      k_{\infty}\sqrt{\frac{(r+\hat{q}_{+})}{(r+\hat{q}_{-})}}
      \left[
      1-\frac{\alpha'}{8}(\beta_{+}\beta_{-}+1)\frac{\hat{q}_{+}\hat{q}_{-}}{
      (r+\hat{q}_{+})(r+\hat{q}_{-})(r+\hat{q}_{0})(r+\hat{q})}
      \right]\,,
    \\
    & \nonumber \\
    k^{(1)}
    & =
      k_{\infty}\sqrt{\frac{(r+\hat{q}_{+})}{(r+\hat{q}_{-})}}
      \left[
      1+\frac{\alpha'}{8}(\beta_{+}\beta_{-}+1)\frac{\hat{q}_{+}\hat{q}_{-}}{
      (r+\hat{q}_{+})(r+\hat{q}_{-})(r+\hat{q}_{0})(r+\hat{q})}
      \right]\,.
  \end{align}
\end{subequations}

T~duality in the $w$ direction interchanges the 1-forms $A^{w}$ and
$B^{(1)}{}_{w}$ and the scalar combinations $\ell$ and $\ell^{(1)\, -1}$ and
it is equivalent to the following transformations in the parameters of the
solution:

\begin{equation}
  T_{w}:
  \,\,\,\,\,
  \beta_{0}
  \leftrightharpoons
  \beta\,,
  \,\,\,\,\,
  \hat{q}_{0}
    \leftrightharpoons
  \hat{q}\,,
  \,\,\,\,\,
  \ell_{\infty}
    \leftrightharpoons  
  \ell_{\infty}^{-1}\,.
\end{equation}

T~duality in the $z$ direction interchanges the 1-forms $A^{z}$ and
$B^{(1)}{}_{z}$ and the scalar combinations $k$ and $k^{(1)\, -1}$ and
it is equivalent to the following transformations in the parameters of the
solution:

\begin{equation}
  T_{z}:
  \,\,\,\,\,
  \beta_{+}
      \leftrightharpoons
  \beta_{-}\,,
  \,\,\,\,\,
  \hat{q}_{+}
      \leftrightharpoons
  \hat{q}_{-}\,,
  \,\,\,\,\,
  k_{\infty}
    \leftrightharpoons  
  k_{\infty}^{-1}\,.
\end{equation}
The 4-dimensional dilaton and the metric function $e^{2U}$ are manifestly
invariant under these transformations

\begin{subequations}
  \begin{align}
    e^{-2\phi}
    & =
    e^{-2\phi_{\infty}}\sqrt{\frac{(r+\hat{q}_{+})(r+\hat{q}_{-})}{(r+\hat{q}_{0})(r+\hat{q})}}
    \left\{1
      -\frac{\alpha'}{8}\left[\frac{\hat{q}^{\,2}(r+\hat{q}_{0})^{2}+\hat{q}_{0}^{2}(r+\hat{q})^{2}}{(r+\hat{q}_{0})^{3}(r+\hat{q})^{3}}
      \right.\right.
      \nonumber 
        \\
        & \nonumber \\
        & \hspace{.5cm}
        \left.\left.
        +(\beta_{+}\beta_{-}+1)\frac{\hat{q}_{+}\hat{q}_{-}}{
          (r+\hat{q}_{+})(r+\hat{q}_{-})(r+\hat{q}_{0})(r+\hat{q})}\right]\right\}\,,
        \\
    & \nonumber \\
    e^{2U}
    & =
     \frac{1}{r^{2}} \sqrt{(r+\hat{q}_{+})(r+\hat{q}_{-})(r+\hat{q}_{0})(r+\hat{q})}
    \left\{1
      +\frac{\alpha'}{8}\left[\frac{\hat{q}^{\,2}(r+\hat{q}_{0})^{2}+\hat{q}_{0}^{2}(r+\hat{q})^{2}}{(r+\hat{q}_{0})^{3}(r+\hat{q})^{3}}
      \right.\right.
      \nonumber 
        \\
        & \nonumber \\
        & \hspace{.5cm}
        \left.\left.
        -(\beta_{+}\beta_{-}+1)\frac{\hat{q}_{+}\hat{q}_{-}}{
          (r+\hat{q}_{+})(r+\hat{q}_{-})(r+\hat{q}_{0})(r+\hat{q})}\right]\right\}\,.
  \end{align}
\end{subequations}
In terms of the $\hat{q}$ charge parameters, the mass and the area of the
horizon are T-duality invariant:\footnote{One should remember that one of the
  boundary conditions imposed to determine the $\alpha'$ corrections was that
  the near-horizon behaviour of the $\mathcal{Z}_{\pm,0,0h},\mathcal{H}$
  functions (and, hence, of the solution) is not modified. Therefore, the coefficients of the
  their $1/r$ terms in the $r\rightarrow 0$ limit are the same they were at
  zeroth order, namely the unhatted charge parameters $q$. The area of the
  horizon is naturally given in terms of them. On the other hand, the mass
  depends on the coefficients of the $1/r$ terms in the $r\rightarrow \infty$
  limit, which in this case are the hatted charge parameters.}

\begin{subequations}
  \begin{align}
    M
    & =
      \frac{1}{4G_{N}^{(4)}}\left(\hat{q}_{+}+\hat{q}_{-}+\hat{q}_{0}+\hat{q}\right)\,,
    \\
    & \nonumber \\
        \label{eq:Abeta0=beta}
    A
    & =
      4\pi \sqrt{q_{+}q_{-}q_{0}q}
    \nonumber \\
    & \nonumber \\
    & =
      4\pi \sqrt{\hat{q}_{+}\hat{q}_{-}\hat{q}_{0}\,\hat{q}}
      \left\{1-\frac{\alpha'}{8}\frac{\beta_{+}\beta_{-}-1}{\hat{q}_{0}\hat{q}}\right\}\,.  
  \end{align}
\end{subequations}

\noindent
We can define, in analogy with the zeroth-order case in Eq.~(\ref{eq:charges4dcase0}), the asymptotic
charges\footnote{Note that in the case of the KK monopole charge
  $\hat{\mathcal{Q}}$, we have $q$ in the right-hand-side, not $\hat{q}$. This
  is because $\hat{\mathcal{Q}}$ is a magnetic charge, and its definition is
  unrelated to the asymptotic behavior of $\mathcal{H}$.}

\begin{subequations}
  \label{eq:charges4dcase1}
\begin{align}
  \hat{\mathcal{Q}}_{+}
  & \equiv
  \frac{g_{s}^{(4)}\ell_{s}}{16\pi G_{N}^{(4)}} \int_{S^{2}_{\infty}}
  e^{-2\phi}k^{2}\star F^{z}
  =
  \beta _{+} \frac{2   \ell_{\infty} k_{\infty}^{2}}{\ell_{s}g_{s}^{2}} \hat{q}_{+}\,,
  \\
  &\nonumber \\
  \hat{\mathcal{Q}}_{-}
  & \equiv 
  \frac{g_{s}^{(4)}\ell_{s}}{16\pi G_{N}^{(4)}} \int_{S^{2}_{\infty}}
  e^{-2\phi}k^{-2}\star G^{(1)}{}_{z}
  =
 \beta _{-} \frac{2  \ell_{\infty}}{\ell_{s} g_{s}^{2}}\hat{q}_{-}\,,
  \\
  &\nonumber \\
  \hat{\mathcal{Q}}_{0}
  & \equiv
  \frac{g_{s}^{(4)}\ell_{s}}{16\pi G_{N}^{(4)}} \int_{S^{2}_{\infty}} G^{(1)}{}_{w}
  =
   \beta_{0}\frac{2 \ell_{\infty}}{\ell_{s}}\hat{q}_{0h}\,,
  \\
  & \nonumber\\
  \hat{\mathcal{Q}}
  & \equiv
  \frac{g_{s}^{(4)}\ell_{s}}{16\pi G_{N}^{(4)}} \int_{S^{2}_{\infty}} F^{w}
  =
  \beta \frac{2}{\ell_{\infty}\ell_{s}} q\,,
\end{align}
\end{subequations}

\noindent
The interpretation of these quantities in the presence of $\alpha'$ corrections is more involved and deserves further
attention. We will nevertheless use these parameters to compare with previous
results for the entropy, as this matches the definition of the
charges often used in the literature 
\cite{Sahoo:2006pm,Sen:2007qy,Prester:2008iu,Faedo:2019xii, Cano:2021dyy}. In terms of these, the mass and the area of the horizon of these solutions
are given by

\begin{subequations}
  \begin{align}
  M
  & =
  \frac{k_{\infty}}{\ell_s}|\hat{\mathcal{Q}}_{-}|+\frac{1}{k_{\infty}\ell_{s}}|\hat{\mathcal{Q}}_{+}|
  +\frac{k_{\infty}}{g_{s}^{2}\ell_{s}}|\hat{\mathcal{Q}}_{0}|
    +\frac{\ell_{\infty}^{2}k_{\infty}}{g_{s}^{2}\ell_{s}}|\hat{\mathcal{Q}}|\,,
    \\
    & \nonumber \\
    \label{eq:A4Gbeta0=beta}
    \frac{A}{4G_{N}^{(4)}}
    & =
      2\pi 
      \sqrt{|\hat{\mathcal{Q}}_{+}\hat{\mathcal{Q}}_{-}\hat{\mathcal{Q}}_{0}\hat{\mathcal{Q}}|}
      \left\{1 -\frac{(\beta_{+}\beta_{-}-1)}{2|\hat{\mathcal{Q}}_{0}\hat{\mathcal{Q}}|}\right\}\,.
  \end{align}
 \end{subequations}
 These expressions are invariant under the two T~duality transformations,
 which act on the charges and moduli that occur in it as follows:

 \begin{subequations}
   \label{eq:Tdualityonmoduli}
  \begin{align}
    T_{w}:
    & \,\,\,
  \hat{\mathcal{Q}}_{0}
      \leftrightharpoons
 \hat{\mathcal{Q}}\,,
  \,\,\,\,\,
  \ell_{\infty}
    \leftrightharpoons  
      \ell_{\infty}^{-1}\,,
  \,\,\,\,\,
  g_{s} \leftrightharpoons  
      g_{s}\ell_{\infty}^{-1}\,,
    \\
    & \nonumber \\
    T_{z}:
    & \,\,\,
  \hat{\mathcal{Q}}_{+}
      \leftrightharpoons
  \hat{\mathcal{Q}}_{-}\,,
  \,\,\,\,\,
  k_{\infty}
    \leftrightharpoons  
      k_{\infty}^{-1}\,,
  \,\,\,\,\,
  g_{s} \leftrightharpoons  
      g_{s}k_{\infty}^{-1}\,.
  \end{align}
\end{subequations}

\subsubsection{Case 2:  \texorpdfstring{$\beta=-\beta_{0}$}{beta equal
    minus beta0 }}
\label{sec-d4-corrections-extremal-alpha=-alpha0andqneqq0}

In this case, obtaining the $\alpha'$ corrections to the solution in
(\ref{eq:4-charge10dsolution}) is more complicated, as we find that all of the
functions in that ansatz receive corrections.  Remarkably, it is possible to
explicitly solve the equations of motion and we get the following lengthy but
analytic solution: 

\begin{subequations}
\label{eq:deltaZs4-chargecase2.2}
\begin{align}\notag
\delta \mathcal{Z}_{0}
  & =
    \frac{1}{4(q_{0}-q)^{5}} \Bigg\{
    -\frac{q(q_{0}-q)^{5}}{(r+q)^{3}}
    +\frac{(q_{0}-q)^{4}(q+q_{0})}{(r+q)^{2}}
    +\frac{\left(q-q_{0}\right)^{4}q_{0}}{(r+q_{0})^{2}}
  \nonumber \\
& \nonumber \\
  &
    +\frac{(2 q_{0}-q)(q_{0}-q)^{3}}{r+q_{0}}
    -\frac{(q_{0}-q)^{3} (8q_{0}+q)}{r+q}
    -\frac{2 q_{0}(2 q^{3}+29 q^{2} q_{0}+12 q q_{0}^{2}-3q_{0}^{3})}{r}
    \nonumber \\
  & \nonumber \\
  &
    -\frac{4 q q_{0}^{2} \left(q{}^{2}+3 q_{0} \left(2
    q+q_{0}\right)\right)}{r^{2}}
    +\frac{2 q q_{0}}{(q_{0}-q)r^{3}}
    \bigg\{q q_{0} \log{\left(\frac{q_{0}}{q}\right)}\bigg[3 q r (q+12r)
    \nonumber \\
  & \nonumber \\
  &
    +q_{0} \left[2 \left(q{}^{2}+18 q r+12 r^{2}\right)
    +3 q_{0} \left(4 q+7 r+2q_{0}\right)\right]
    \bigg]
    \nonumber \\
  & \nonumber \\
  &
    -\log{\left(\frac{r+q_{0}}{r+q}\right)}
    \Big[9 q{}^{2}
    r^{3}+ 3 q_{0} q r \left(q{}^{2}+12 q r+4 r^{2}\right)
    \nonumber \\
  & \nonumber \\
  &
    +q_{0}^{2} \big[2
		q{}^{3}+36 q{}^{2} r+24 q r^{2}-r^{3}+3 q q_{0} \left(4 q+7 r+2
		q_{0}\right)\big]\Big]\bigg\}\Bigg\}\,,\\
& \nonumber \\
\delta \mathcal{Z}_{0h}
  & =
    \frac{(q_{0}-q)[q_{0}(r+q) +q(r+q_{0})]}{4(r+q)^{3}(r+q_{0})^{2}}\,, 
\\
  & \nonumber \\
\delta \mathcal{H}
  & =
    \frac{1}{2(q_{0}-q)^{5}}\Bigg\{\frac{q}{r} \left(12 q_{0} q{}^{2}+29
                     q_{0}^{2} q+2 q_{0}^{3}-3 q^{3}\right)
                     \nonumber \\
  & \nonumber \\
  &
    +\frac{2 q_{0} q^{2}}{r^{2}} \left(6 q_{0} q+q_{0}^{2}+3
    q{}^{2}\right)
    -\frac{q_{0}(q_{0}-q)^{3}}{r+q}
    -\frac{3 q(q_{0}-q)^{3}}{r+q_{0}}
    \nonumber \\
  & \nonumber \\
  &
    +\frac{q_{0} q}{r^{3}(q_{0}-q)}
    \bigg[r^{3} \left(12 q_{0} q+9q_{0}^{2}-q^{2}\right)
    \log{\left(\frac{r+q_{0}}{r+q}\right)}
    \nonumber \\
  & \nonumber \\
  &
    +q_{0} q \Big[12 q_{0} \left(3 r^{2}+3 r q+q{}^{2}\right)+q_{0}^{2} (3
    r+2 q)+3 q \left(8 r^{2}+7 r q+2 q{}^{2}\right)\Big]\times
    \nonumber \\
  & \nonumber \\
  &
    \times\left[
    \log{\left(\frac{r+q_{0}}{r+q}\right)}-\log{\left(\frac{q_{0}}{q}\right)}\right]
    \bigg]\Bigg\}\,,
  \\
& \nonumber \\
\delta \mathcal{Z}_{+}
  & =
    \frac{q_{+}}{q_{-}}\delta \mathcal{Z}_{-}
    +(\beta_{+}\beta_{-}+1)\frac{q_{+} \left[
	(r+q_{-})(r+q+q_{0})+qq_{0}\right]}{4qq_{0}(r+q_{-}) (r+q_{0})(r+q)}\,,
  \\
  & \nonumber \\
\delta \mathcal{Z}_{-}
  & =
    \frac{q_{-}}{4r^{3}(q_{0}-q)^{5}}
    \Bigg\{
    8 q_{0}q r^{3}
    \log{\left(\frac{r+q_{0}}{r+q}\right)}
    \nonumber \\
  & \nonumber \\
  &
    +(q_{0}+q) \bigg[2 q_{0} q (q_{0} q-3 r^{2})
    \log{\left(\frac{1+r/q}{1+r/q_{0}}\right)}+r (q_{0}-q)
    \left[r(q_{0}+q) -2q_{0}q\right]\bigg]
    \Bigg\}\,,
\\
  & \nonumber \\
  \label{Cphi2}
C_{\phi}
&  =
           e^{2\hat{\phi}_{\infty}}\left\{1+\frac{\alpha'}{4}\Upsilon\right\}\,,
\end{align}
\end{subequations}

\noindent
where we have defined\footnote{This quantity is proportional to the correction to the attractor value of the dilaton.}

\begin{equation}
  \label{eq:Upsilon}
  \Upsilon
  =
  \frac{1}{(q_{0}-q)^{5}}\left[6q_{0}q(q_{0}+q)\log{\left(\frac{q_{0}}{q}\right)}
  -(q_{0}-q)(q_{0}^{2}+q^{2}+10q_{0}q)\right]\,.
\end{equation}

As usual, in this solution the integration constants have been fixed so that
the $\delta\mathcal{Z}_{\pm,0,0h}$ and $\delta \mathcal{H}$ functions vanish at
infinity and that they do not have a pole at $r=0$.

The main difference between this case and all the preceding cases is that,
now, the corrections make $\mathcal{Z}_{0h}$ different from $\mathcal{Z}_{0}$
and, therefore, the solution depends on five different functions instead of four,
even though it only has four different charge parameters. Also, this is the only case in which $\mathcal{H}$ receives corrections. 
Observe that it is
$\hat{q}_{0h}\equiv q_{0}+\alpha'\delta q_{0h}$ that occurs in
$B^{(1)}{}_{w}$, and not $\hat{q}_{0}\equiv q_{0}+\alpha'\delta q_{0}$. Thus,
the charge parameter which is relevant for T~duality in the $w$ direction is
$\hat{q}_{0h}$, which turns out to be equal to $q_{0}$ because $\delta q_{0h}$
vanishes. On the other hand, it is the charge parameter $q$, that occurs in
$A^{w}$ and is interchanged with $q_{0}$ under T~duality in the $w$
direction. Therefore, we will keep using $q_{0}$ and $q$, but we will define

\begin{subequations}
  \begin{align}
    \label{eq:hatq+beta0=-beta}
    \hat{q}_{+}
    & =
      q_{+}\left\{1
      -\frac{\alpha'}{4}\left[\Upsilon-(\beta_{+}\beta_{-}+1)\frac{1}{q_{0}q}\right]
      \right\}\,,
    \\
    & \nonumber \\
    \label{eq:hatq-beta0=-beta}
    \hat{q}_{-}
    & =
      q_{-}\left\{1-\frac{\alpha'}{4}\Upsilon  \right\}\,,
  \end{align}
\end{subequations}

\noindent
where $\Upsilon$ is given by Eq.~(\ref{eq:Upsilon}). As we will see, these are the changes interchanged by a T duality along the $z$ direction. Thus, it is convenient to rewrite the solution in terms of these charges. The functions $\mathcal{Z}_{+}$ and $\mathcal{Z}_{-}$ are given by

\begin{subequations}
  \begin{align}
    \mathcal{Z}_{+}
    & =
      1 +\frac{\hat{q}_{+}}{r} +\frac{\alpha'}{4}\left[\hat{q}_{+}G(r)
      -(\beta_{+}\beta_{-}+1)
      \frac{\hat{q}_{+}\hat{q}_{-}}{r(r+\hat{q}_{-})(r+\hat{q}_{0})(r+\hat{q})}\right]\,,
    \\
    & \nonumber \\
    \mathcal{Z}_{-}
    & =
      1 +\frac{\hat{q}_{-}}{r} +\frac{\alpha'}{4}\hat{q}_{-}G(r)\,,
  \end{align}
\end{subequations}

\noindent
where we have defined

\begin{equation}
  \begin{aligned}
    G(r)
    & =
    \frac{1}{(q_{0}-q)^{5}r^{3}} \Bigg\{
   8q_{0}q
   r^{3}\left[\log{\left(\frac{r+q_{0}}{r+q}\right)}-\frac{q_{0}-q}{r}\right]
   +2q_{0}^{2}q^{2}(q_{0}+q)\log{\left(\frac{1+r/q}{1+r/q_{0}}\right)}
   \\
    & \\
    & \hspace{.5cm}
    +6q_{0}q(q_{0}+q)r^{2}\log{\left(\frac{r+q_{0}}{r+q}\right)}
    -2q_{0}q(q_{0}^{2}-q^{2})r\Bigg\}\,.
  \end{aligned}
\end{equation}

\noindent
The rest of the functions of the solution do not change, because they depend
on $q_{0}$ and $q$, only. Nevertheless, let us quote the value of the
corrections to the coefficients of $1/r$ in the the expansions of
$\mathcal{Z}_{0}$ and $\mathcal{H}$ at spatial infinity, $\delta q_{0}$ and
$\delta{q}$ respectively. They are defined defined by

\begin{subequations}
  \begin{align}
    \mathcal{Z}_{0}
    & \stackrel{\rho\rightarrow \infty}{\sim}
      1 +\frac{q_{0}+\alpha'\delta q_{0}}{r}
      +\mathcal{O}\left(\frac{1}{r^{2}} \right)\,,
    \\
    & \nonumber \\
    \mathcal{H}
    &  \stackrel{\rho\rightarrow \infty}{\sim}
      1 +\frac{q+\alpha'\delta q}{r}
      +\mathcal{O}\left(\frac{1}{r^{2}} \right)\,,
  \end{align}
\end{subequations}

\noindent
and take the form

\begin{subequations}
  \begin{align}
    \label{eq:deltaq0}
    \delta q_{0}
    & =
      \frac{1}{2(q_{0}-q)^{5}}
      \left\{ q^{4} -11q_{0}q^{3}-47 q_{0}^{2}q^{2} -3q_{0}^{3}q
      +12\frac{q_{0}q}{(q_{0}-q)}\log{\left(q_{0}/q\right)}
      \right\}\,,
    \\
    & \nonumber \\
    \label{eq:deltaq}
    \delta q
    & =
      -\frac{1}{2(q_{0}-q)^{5}}
      \left\{ q_{0}^{4} -11q_{0}^{3}q-47 q_{0}^{2}q^{2} -3q_{0}q^{3}
      -12\frac{q_{0}q}{(q_{0}-q)}\log{\left(q_{0}/q\right)}
      \right\}\,.
  \end{align}
\end{subequations}

\noindent
These quantities determine the correction to the mass of the solution, which will be computed in Section~\ref{sec-d4-mass}, but we
remark that $q_{0}+\alpha'\delta q_{0}$ and  $q+\alpha'\delta q$ are not charges (the asymptotic charges are $\hat{q}_{0h}=q_0$ and $q$, as we stressed earlier), 

The Kaluza-Klein and winding vectors associated to the internal $z$ direction
are in this case given by

\begin{subequations}
  \begin{align}
    A^{z}{}_{\mu}
    & =
      \beta_{+}k_{\infty}^{-1}\left\{-1+\frac{r}{r+\hat{q}_{+}}\left[1
      -\frac{\alpha'}{4}\left[\frac{\hat{q}_{+}r}{r+\hat{q}_{+}}G(r)
      +\frac{(\beta_{+}\beta_{-}+1)\hat{q}_{+}\hat{q}_{-}}{(r+\hat{q}_{+})(r+\hat{q}_{-})(r+q_{0})(r+q)}\right]\right]\right\}\,,
    \\
    & \nonumber \\
    B^{(1)}{}_{z\, \mu}
    & =
      \beta_{-}k_{\infty}\left\{-1+\frac{r}{r+\hat{q}_{-}}\left[1
      -\frac{\alpha'}{4}\left[\frac{\hat{q}_{-}r}{r+\hat{q}_{-}}G(r)
      +\frac{(\beta_{+}\beta_{-}+1)\hat{q}_{+}\hat{q}_{-}}{(r+\hat{q}_{+})(r+\hat{q}_{-})(r+q_{0})(r+q)}\right]\right]\right\}\,,
  \end{align}
\end{subequations}

\noindent
while the Kaluza-Klein scalar $k$ and the scalar combination $k^{(1)}$ are
given by

\begin{subequations}
  \begin{align}
    k
    & =
      k_{\infty}\sqrt{\frac{r+\hat{q}_{+}}{r+\hat{q}_{-}}}
      \left\{1+\frac{\alpha'}{8}\left[\frac{r^{2}(\hat{q}_{+}-\hat{q}_{-})}{(r+\hat{q}_{+})(r+\hat{q}_{-})}G(r)
      -\frac{(\beta_{+}\beta_{-}+1)\hat{q}_{+}\hat{q}_{-}}{(r+\hat{q}_{+})(r+\hat{q}_{-})(r+q_{0})(r+q)}\right]
      \right\}\,,
    \\
    & \nonumber \\
    k^{(1)}
    & =
      k_{\infty}\sqrt{\frac{r+\hat{q}_{+}}{r+\hat{q}_{-}}}
      \left\{1+\frac{\alpha'}{8}\left[\frac{r^{2}(\hat{q}_{+}-\hat{q}_{-})}{(r+\hat{q}_{+})(r+\hat{q}_{-})}G(r)
      +\frac{(\beta_{+}\beta_{-}+1)\hat{q}_{+}\hat{q}_{-}}{(r+\hat{q}_{+})(r+\hat{q}_{-})(r+q_{0})(r+q)}\right]
      \right\}\,.
  \end{align}
\end{subequations}

\noindent
The dilaton and metric function are given by

\begin{subequations}
  \begin{align}
    e^{2\phi}
    & =
      e^{2\phi_{\infty}}
      \sqrt{\frac{(r+q_{0})(r+q)}{(r+\hat{q}_{+})(r+\hat{q}_{-})}}
      \left\{
      1+\frac{\alpha'}{2}
      \left[
\frac{\delta\mathcal{Z}_{0}}{\mathcal{Z}_{0}}
+\frac{\delta\mathcal{H}}{\mathcal{H}}
      -\tfrac{1}{2}r^{2}G(r)'
      \right.\right.
      \nonumber \\
    & \nonumber \\
    & \hspace{.5cm}
      \left.\left.
      +\frac{(\hat{q}_{+}+\hat{q}_{-})r^{2}+2\hat{q}_{+}\hat{q}_{-}r}{4(r+\hat{q}_{+})(r+\hat{q}_{-})}G(r)
      +\frac{(\beta_{+}\beta_{-}+1)\hat{q}_{+}\hat{q}_{-}}{4(r+\hat{q}_{+})(r+\hat{q}_{-})(r+q_{0})(r+q)}
      \right]
      \right\}\,,
    \\
    & \nonumber \\
        Fe^{2U}
    & =
     \frac{1}{r^{2}} \sqrt{(r+\hat{q}_{+})(r+\hat{q}_{-})(r+q_{0})(r+q)}
      \left\{
      1+\frac{\alpha'}{2}
      \left[
\frac{\delta\mathcal{Z}_{0}}{\mathcal{Z}_{0}}
+\frac{\delta\mathcal{H}}{\mathcal{H}}
      -\tfrac{1}{2}r^{2}G(r)'
      \right.\right.
      \nonumber \\
    & \nonumber \\
    & \hspace{.5cm}
      \left.\left.
      -\frac{(\hat{q}_{+}+\hat{q}_{-})r^{2}+2\hat{q}_{+}\hat{q}_{-}r}{4(r+\hat{q}_{+})(r+\hat{q}_{-})}G(r)
      -\frac{(\beta_{+}\beta_{-}+1)\hat{q}_{+}\hat{q}_{-}}{4(r+\hat{q}_{+})(r+\hat{q}_{-})(r+q_{0})(r+q)}
      \right]
      \right\}\,,
  \end{align}
\end{subequations}

\noindent
and the Kaluza-Klein scalar $\ell$ and the scalar combination $\ell^{(1)}$ are
given by

\begin{subequations}
  \begin{align}
  \ell
    & =
      \ell_{\infty}\sqrt{\frac{(r+\hat{q}_{0})}{(r+\hat{q})}}
      \left\{
      1+\frac{\alpha'}{2}\left[
      \frac{\delta \mathcal{Z}_{0}}{\mathcal{Z}_{0}}
      - \frac{\delta \mathcal{H}}{\mathcal{H}}
      \right]\right\}\,,
    \\
    & \nonumber \\
    \ell^{(1)}
    & =
      \ell_{\infty}\sqrt{\frac{(r+q_{0})}{(r+q)}}
      \left\{1+\frac{\alpha'}{2}\left[
            \frac{\delta \mathcal{Z}_{0}}{\mathcal{Z}_{0}}
      - \frac{\delta \mathcal{H}}{\mathcal{H}}
      -\frac{\left[q_{0}(r+q)-q(r+q_{0})\right]^{2}}{2(r+q_{0})^{3}(r+q)^{3}}
\right]\right\}\,.
  \end{align}
\end{subequations}

It can be seen that T~dualities along the $w$ and $z$ directions are equivalent to the transformation of parameters:

\begin{subequations}
  \begin{align}
\label{eq:Tdualityw}
    T_{w}:
    & \,\,\,
      \beta_{0} \leftrightharpoons \beta \,,
      \hspace{.5cm}
      q_{0} \leftrightharpoons q \,,
      \hspace{.5cm}
      \ell_{\infty} \leftrightharpoons \ell_{\infty}^{-1}\,,
    \\
    & \nonumber \\
\label{eq:Tdualityz}
    T_{z}:
    & \,\,\,
      \beta_{+} \leftrightharpoons \beta_{-} \,,
      \hspace{.5cm}
      \hat{q}_{+} \leftrightharpoons \hat{q}_{-} \,,
      \hspace{.5cm}
      k_{\infty} \leftrightharpoons k_{\infty}^{-1}\,.
  \end{align}
\end{subequations}

\subsubsection{Corrections to the extremal mass and area of the horizon for $\beta_{0}=-\beta$}
\label{sec-d4-mass}

A very interesting property of the non-supersymmetric solution with
$\beta_{0}=-\beta$ is that the function $\mathcal{H}$ gets $\alpha'$ corrections, while the corrections to ${\cal Z}_{0}$ are different from those of ${\cal Z}_{0h}$. This in turn implies that, unlike for the rest of the
4-dimensional and 5-dimensional black hole solutions we have studied in this
paper, the mass of these non-supersymmetric black holes receives a non-trivial
correction when expressed in terms of the charges. From the
asymptotic behavior of the $g_{tt}$ component of the metric, 

\begin{equation}
  g_{tt}
  \sim
  1-\tfrac{1}{2}\left[\hat{q}_{+}+\hat{q}_{-}
    +(q_{0}+\alpha'\delta q_{0}) +(q+\alpha'\delta q)\right]\frac{1}{r}
  +\mathcal{O}\left(\frac{1}{r^{2}}\right)\,,
\end{equation}

\noindent
we obtain

\begin{equation}
  \begin{aligned}
    M
    & =
    \frac{1}{4G_{N}^{(4)}}\left\{\hat{q}_{+}+\hat{q}_{-}+q_{0}+q
    \right.
    \\
    & \\
    & \hspace{.5cm}
    \left.
      -\frac{\alpha'}{2(q_{0}-q)^{5}} \left[
        (q_{0}^{2}-q^{2})(q_{0}^{2}+q^{2}-8q_{0}q)
        +12q_{0}^{2}q^{2}\log{\left(q_{0}/q\right)} \right] \right\}\,,
  \end{aligned}
\end{equation}

\noindent
after making use of Eqs.~(\ref{eq:deltaq0})
and (\ref{eq:deltaq}). It is straightforward to check that the mass is invariant under both T dualities \eqref{eq:Tdualityw} and \eqref{eq:Tdualityz}.

Let us now rewrite the mass in terms of the charges  defined in Eqs.~(\ref{eq:charges4dcase1}). The expression of these charges in terms of the ${\hat q}_{+}$, ${\hat q}_{-}$, ${\hat q}_{0h}$ and $q$ does not change. However, one has to bear in mind that the relation between the latter and $q_{+}$, $q_{-}$ and $q_{0}$ is different from the $\beta_{0}=\beta$ case. Taking this into account, we obtain



\begin{equation}
  M
  =
  \frac{k_{\infty}}{\ell_s}|\hat{\mathcal{Q}}_{-}|
  +\frac{1}{\ell_{s}k_{\infty}}|\hat{\mathcal{Q}}_{+}|
  +\frac{k_{\infty}}{\ell_{s}g_{s}^{2}}|\hat{\mathcal{Q}}_{0}|
  +\frac{\ell_{\infty}^{2}k_{\infty}}{\ell_{s}g_{s}^{2}}|\hat{\mathcal{Q}}|+\delta M \,,
\end{equation}

\noindent
where the mass shift reads

\begin{equation}
\label{eq:shiftM4d}
\delta M
=
-\frac{16\ell_{\infty}^{2} k_{\infty}}{\ell_{s}g_{s}^{2} |\hat{\mathcal{Q}}_{0}|}f(v)\,,
\end{equation}

\noindent
and $f(v)$ is 

\begin{equation}\label{eq:fv}
  f(v)
  =
  \frac{v^{4}-8 v^{3}+8 v-1+12 v^{2} \log{\left(v\right)}}{(v-1)^{5}}\,,
  \quad
  v
  =
  \frac{|\hat{\mathcal{Q}}|}{|\hat{\mathcal{Q}}_{0}|}\ell_{\infty}^{2}\,.
\end{equation}

\begin{figure}[t!]
	\begin{center}
		\includegraphics[width=0.55\textwidth]{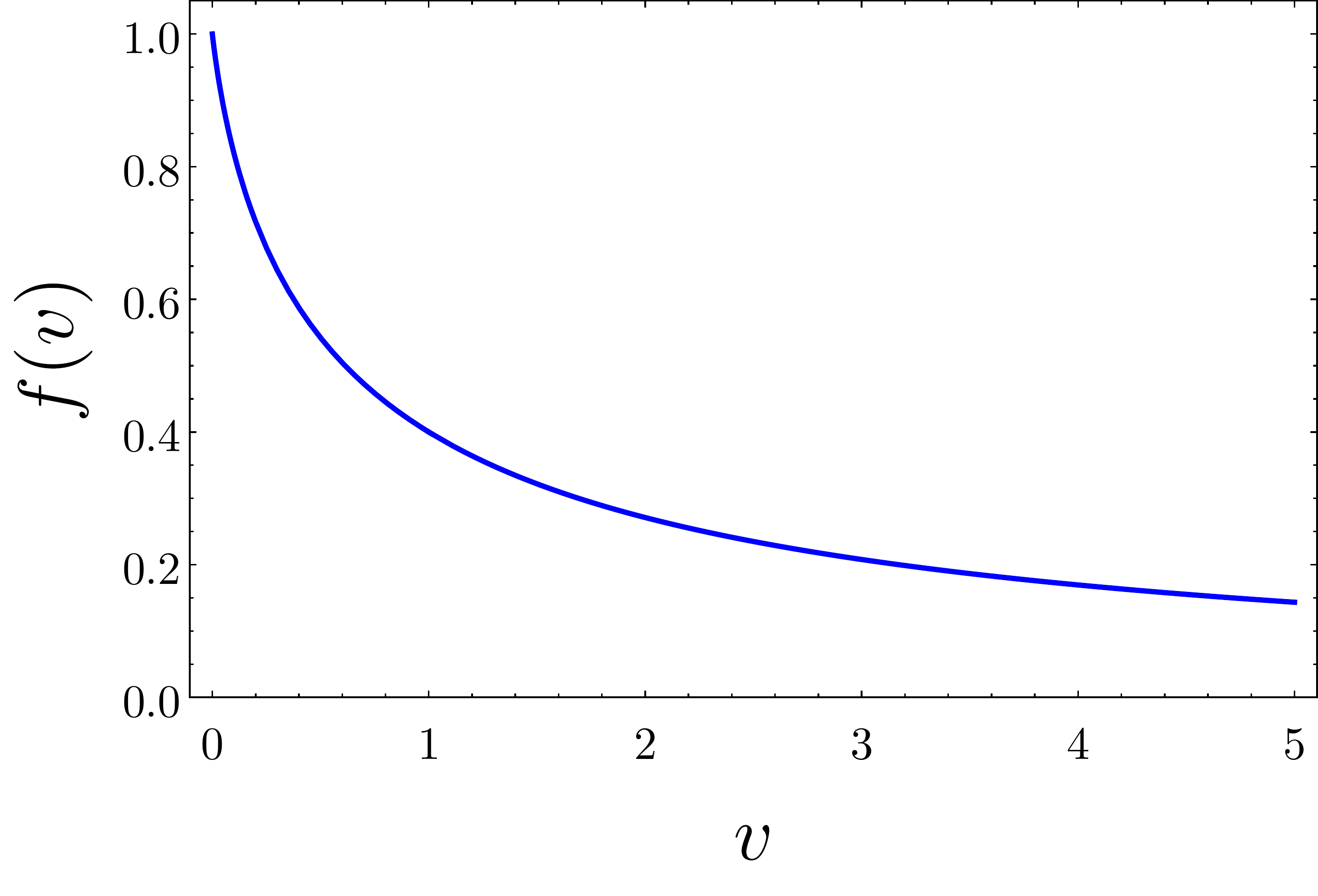} 
		\caption{Function $f(v)$ defined in (\ref{eq:fv}) that
                  controls the shift to the mass (\ref{eq:shiftM4d}). Observe
                  that it is smooth everywhere and that it is positive,
                  meaning that the mass is corrected negatively, in agreement
                  with the mild form of the WGC.}
		\label{fig:fv}
	\end{center}
\end{figure}

This result deserves further comments. First of all, let us note that, despite
the apparent singularity of $f(v)$ for $v=1$, this function is actually smooth
for every $v\ge 0$, as illustrated in Fig.~\ref{fig:fv}. In fact, we have $\lim_{v\rightarrow 1} f(v)=2/5$. Second,
the higher-derivative corrections to the mass of extremal black holes are a
matter of interest in the context of the weak gravity conjecture
\cite{Cheung:2018cwt,Hamada:2018dde,Bellazzini:2019xts,Charles:2019qqt,Loges:2019jzs,Cano:2019oma,Cano:2019ycn,Andriolo:2020lul,Loges:2020trf,Cano:2020qhy,Cano:2021tfs,Arkani-Hamed:2021ajd}. Originally,
this conjecture has been formulated for black holes charged under a single
$U(1)$ field, and it states that, in a consistent theory of quantum gravity,
the corrections to the extremal charge-to-mass ratio $Q/M$ should be
positive. The logic of this statement lies in the fact that, in this way, the
decay of extremal black holes is possible in terms of energy and charge
conservation.  The extension of this conjecture to the case of black holes
with multiple charges is subtle \cite{Jones:2019nev}, but as a general rule,
one can see that the corrections to the mass should be negative in order to
allow for the decay of extremal black holes. Now, since our black holes are an
explicit solution of string theory, they should satisfy the WGC, assuming it
is correct. We check that, indeed, $f(v)>0$ for every $v\ge 0$ (see
Fig.~\ref{fig:fv}), which implies that $\delta M<0$ for all the values of the
charges.  Thus, this is a quite non-trivial test of the validity of the WGC in
string theory, that adds up to the ones already found in
Refs.~\cite{Cano:2019oma,Cano:2019ycn}.

%
%
%

Finally, the area of the event horizon is easily found to be 

\begin{equation}
    \label{eq:Abeta0=-beta}
  A = 4\pi \sqrt{q_{+}q_{-}q_{0}q}
  \left\{1-\frac{\alpha'}{4}\Upsilon\right\}\,,  
\end{equation}

\noindent
where $\Upsilon$ is given in Eq.~(\ref{eq:Upsilon}). This expression is also
invariant under the T~duality interchanges
$\hat{q}_{+}\leftrightharpoons \hat{q}_{-}$ and $q_{0} \leftrightharpoons
q$. In terms of the charges defined in \eqref{eq:charges4dcase1}, it takes the form

\begin{equation}
  \label{eq:A4Gbeta0=-beta}
  \frac{A}{4G_{N}^{(4)}}
  =
  2\pi
\sqrt{|\hat{\mathcal{Q}}_{+}\hat{\mathcal{Q}}_{-}\hat{\mathcal{Q}}_{0}\hat{\mathcal{Q}}|}\left\{1-\frac{\pi(\beta_{+}\beta_{-}+1)}{2|\hat{\mathcal{Q}}_{0}\hat{Q}|}\right\}\,.
\end{equation}

\subsubsection{Case 2-1:  \texorpdfstring{$\beta=-\beta_{0}$}{beta equal
    minus beta0 } and \texorpdfstring{$q=q_{0}$}{q equal
    q0 }}
\label{sec-d4-corrections-extremal-alpha=-alpha0andqeqq0}

It is worth mentioning that, despite the apparent singularity of the previous
solution at $q=q_{0}$, the the limit $q\rightarrow q_{0}$ is well-defined, and
the solution simplifies greatly in that case:\footnote{A useful result is
  \begin{equation}
  \lim_{q\rightarrow q_{0}}\Upsilon = -\frac{1}{10 q_{0}^{2}}\,.  
  \end{equation}
}

\begin{subequations}
\label{eq:deltaZs4-chargecase2.1}
\begin{align}
\delta \mathcal{Z}_{0}
& = 
-\frac{1}{120 q_{0}(r+q_{0})^{3}}\left(12r^{2}+45q_{0}r +65q_{0}^{2}\right)\,, 
\\
& \nonumber \\
\delta \mathcal{Z}_{0h}
& = 
0\,, 
\\
& \nonumber \\
\delta \mathcal{H}
& = 
\delta\mathcal{Z}_{0}\,, 
\\
& \nonumber \\
\delta \mathcal{Z}_{+}
  & =
    \frac{q_{+}}{q_{-}}\delta
    \mathcal{Z}_{-}+(\beta_{+}\beta_{-}+1)\frac{q_{+}
    \left[(r+q_{-})(2r+q_{0})+q_{0}^{2}\right)}{4q_{0}^{2}(r+q_{-})(r+q_{0})^{2}}\,,
  \\
& \nonumber \\
\delta \mathcal{Z}_{-}
& = 
\frac{q_{-}}{120 q_{0}^{2}(r+q_{0})^{3}}\left(8r^{2}+9q_{0}r +3q_{0}^{2}\right)\,, 
\\
  & \nonumber \\
C_{\phi}
&  =
e^{2\hat{\phi}_{\infty}}\left(1-\frac{\alpha'}{40q_{0}^{2}}\right)\,.
\end{align}
\end{subequations}

The mass and the area of the horizon are given by

\begin{subequations}
  \begin{align}
    M
    & =
    \frac{1}{4G_{N}^{(4)}}\left\{\hat{q}_{+}+\hat{q}_{-}+2q_{0}
      -\frac{\alpha'}{5q_{0}}\right\}\,,
    \\
    & \nonumber \\
    A & =
        4\pi \sqrt{\hat{q}_{+}\hat{q}_{-}q_{0}q}\left\{1-\frac{\alpha'}{40q_{0}^{2}}\right\}\,.
  \end{align}
\end{subequations}

\subsection{$\alpha'$-corrected entropy}
\label{sec-d4-entropy}

The computation of the black hole entropy parallels that of section
\ref{sec-d5-entropy}.  From the equation
(\ref{eq:Waldentropyformula}), one can see that the Wald entropy can be expressed
as

\begin{equation}
  \label{eq:entropy4d1}
  S
  =
  \frac{A}{4G_{N}^{(4)}}+\Delta S\,.
\end{equation}

\noindent
The area term $A/(4G_{N}^{(4)})$ has been computed in Eqs.~(\ref{eq:A4Gbeta0=beta}) and
(\ref{eq:A4Gbeta0=-beta}) for the $\beta =\beta_{0}$ and $\beta=-\beta_{0}$
cases, respectively. In terms of the original charge parameters, it is given
by 
\begin{equation}
A=\frac{4\pi e^{2{\hat\phi}_{\infty}}}{C_{\phi}}\sqrt{q_{+}q_{-}q_0q}\, ,
\end{equation}

\noindent
where the constant $C_{\phi}$ is given by (\ref{Cphi1}) and (\ref{Cphi2}) in
each case. 



\noindent
On the other hand, $\Delta S$, the correction to the Bekenstein-Hawking
entropy, is given by

\begin{equation}
  \label{eq:eq:entropy4d2}
  \Delta S
  =
  -\frac{\alpha'}{16G_{N}^{(10)}}
 \int_{\Sigma}
 e^{-2(\hat\phi-{\hat\phi}_{\infty})}\hat\star {\hat H}^{(0)} \wedge {\hat\Omega}^{(0)}_{(-)}{}^{ab}{\hat n}_{ab}\,,
\end{equation}

\noindent
where the integral is taken at $r=0$ and where, in the Zehnbein basis
in Eqs.~(\ref{eq:basis4d}), the only non-vanishing component of the binormal is
${\hat n}^{01}=-{\hat n}_{01}=+1$.  It is important to note that, in the limit
$r\rightarrow 0$, we have

\begin{align}
  \lim_{r\rightarrow 0}{\hat \Omega}^{(0)}_{(-)}{}^{01}
  & =
    \frac{1}{\sqrt{q_{0}q}}{\hat e}^{t}
    -\frac{\left(\beta_{+}+\beta_{-}\right)}{\sqrt{q_{0}q}}{\hat e}^{z}\,,
  \\
  & \nonumber \\
  \lim_{r\rightarrow 0}e^{-2(\hat\phi-{\hat\phi}_{\infty})}
  & =
    \frac{q_{-}}{q_{0}}e^{2\hat\phi_{\infty}}C_{\phi}^{-1}\,.
\end{align}

\noindent
Integrating over the angular and compact variables we obtain that, for both cases
$\beta_{0}\beta=\pm 1$, $\Delta S$ is given by

\begin{equation}
  \Delta S
  =
  \frac{\alpha'\pi e^{2{\hat \phi}_{\infty}}}{4C_{\phi}G_{N}^{(4)}}
  \sqrt{\frac{q_{+}q_{-}}{q_{0}q}}(\beta_{+}\beta_{-}+1)
  \,.
\end{equation}
Putting together both contributions, we find that the entropy reads

\begin{equation}
  \label{entropysmallq}
  S
  =
  \frac{\pi e^{2\hat\phi_{\infty}}}{C_{\phi}G_{N}^{(4)}}
  \sqrt{q_{+}q_{-}q_0q}\left(1+\frac{\alpha'(\beta_{+}\beta_{-}+1)}{4q_0q}\right)\,,
\end{equation}

\noindent
and we need to express it in terms of the charges $\hat{\mathcal{Q}}_i$, which
are defined by Eq.~(\ref{eq:charges4dcase1}).  The relations between the
charges and the $q_i$ parameters can be written jointly for the
$\beta_0\beta=+1$ and $\beta_0\beta=-1$ cases as

\begin{equation}
  \label{eq:charges4dcase3}
\begin{aligned}
  \hat{\mathcal{Q}}_{+}
  &=
  \beta _{+} \frac{2   \ell_{\infty} k_{\infty}^{2}}{\ell_{s} C_{\phi}} q_{+}\left[1+\frac{\alpha'(\beta_{+}\beta_{-}+1)}{4q_0q}\right]\,,
  \\
  & \\
  \hat{\mathcal{Q}}_{-}  &=
 \beta _{-} \frac{2  \ell_{\infty}}{\ell_{s} C_{\phi}}q_{-}\,,
  \\
  & \\
  \hat{\mathcal{Q}}_{0}  &=
   \beta_{0}\frac{2 \ell_{\infty}}{\ell_{s}}q_0\left[1-\frac{\alpha'(\beta_{0}\beta+1)}{4q_0q}\right]\,,
  \\
  & \\
  \hat{\mathcal{Q}}
  &=
  \beta \frac{2}{\ell_{\infty}\ell_{s}} q\,,
\end{aligned}
\end{equation}

\noindent
Inserting these relations in Eq.~(\ref{entropysmallq}), we find the following
expression for the entropy,\footnote{Interestingly, there is no approximation implied from Eq.~(\ref{entropysmallq}) to Eq.~(\ref{eq:entropy4dq}).}

\begin{equation}
 \label{eq:entropy4dq}
  S
  =
  2\pi\sqrt{|\hat{\mathcal{Q}}_{+}\hat{\mathcal{Q}}_{-}|(|\hat{\mathcal{Q}}_{0}\hat{\mathcal{Q}}|+2+\beta_{+}\beta_{-}+\beta_{0}\beta)}\,.
\end{equation}

Let us comment on this result. First, observe that the entropy is trivially
invariant under the T~duality transformations
$\hat{\mathcal{Q}}_{+}\leftrightharpoons \hat{\mathcal{Q}}_{-}$ or
$\hat{\mathcal{Q}}_{0}\leftrightharpoons\hat{\mathcal{Q}}$.  Second, note that
in the supersymmetric case $\beta_{+}\beta_{-}=\beta_{0}\beta=1$, our result
matches \emph{exactly} the microscopic value of the entropy that can be
obtained from \cite{Kutasov:1998zh}, which is supposed to be
$\alpha'$-exact. In fact, we reproduce the famous $+4$ term of
\cite{Kutasov:1998zh} inside the square root.  This result also matches those
of Refs.~\cite{Sahoo:2006pm,Sen:2007qy,Prester:2008iu,Faedo:2019xii, Cano:2021dyy} obtained
from the application of the entropy function method and the one in
\cite{Ortin:2020xdm} from the application of the Lorentz-invariant Wald
formula (\ref{eq:Waldentropyformula}). The supersymmetry-breaking case
$\beta_{+}\beta_{-}=-1$, $\beta_{0}\beta=1$ has also been studied in the
literature by means of the entropy function method, and our result matches the
$+2$ factor of Ref.~\cite{Sen:2007qy} (see also
\cite{Sahoo:2006pm,Prester:2008iu}). Finally the cases with
$\beta_{0}\beta=-1$ do not seem to have been studied in the literature, but we
see that when $\beta_{+}\beta_{-}=1$ we get again the $+2$ factor, while if we
break supersymmetry twice, $\beta_{+}\beta_{-}=-1$ , then there is no
correction to the zeroth-order entropy. Thus, the more ways we break
supersymmetry, the smaller the entropy. For a fixed value of the absolute
values of the charges, the supersymmetric solution is the most entropic one.

\section{$\alpha'$-corrected T~duality}
\label{sec-Tduality}

During the last three decades, the invariance of the zeroth-order in $\alpha'$
string effective action under Buscher's T~duality transformations
\cite{Buscher:1987sk,Buscher:1987qj} and its type~II generalizations
\cite{Bergshoeff:1995as} has been profusely used as a solution-generating
technique. In contrast, the generalization of the rules to first order in
$\alpha'$ in the context of the heterotic superstring effective action
\cite{Bergshoeff:1995cg} (see also \cite{Elgood:2020xwu}) has not, essentially
because almost no solutions to the first-order equations had been found until
the $\alpha'$ corrections to the Strominger-Vafa black hole were computed in
Ref.~\cite{Cano:2018qev}. The first-order in $\alpha'$ T~duality rules of
Ref.~\cite{Bergshoeff:1995cg} were tested in this solution as well as in the
4-dimensional 4-charge solution of Ref.~\cite{Cano:2018brq}. Since these
solutions, as a family, are expected to be invariant under T~duality (only the
charges and moduli are transformed, while the functional form of the solution
remains invariant) T~duality provided a test of the transformations and of the
corrections.

The solutions considered in this paper are various non-supersymmetric
generalizations of the solutions considered in
Refs.~\cite{Cano:2018qev,Cano:2018brq}. They are expected to be invariant
under T~duality and this condition can, again, be used to test the corrections
found. We have, indeed, checked that the solutions can be written in
manifestly T~duality-invariant forms in a highly non-trivial way.

Inspired by our results, in this section we are going to show that T~duality
can be used to constrain the form of the first-order in $\alpha'$ corrections
when we do not know them. In some cases, these constraints essentially
determine the corrections. This fact can be used to simplify the problem of
finding the corrections. The key to this result is the fact that the
definitions of some of the lower-dimensional fields that transform nicely
under T~duality (actually, all those descending from the 10-dimensional
Kalb-Ramond 2-form) contain $\alpha'$ corrections while the rest, do not. This
means that the corrections to the latter in the solutions must be related to
the corrections to the definitions of the former. This mechanism will become
clearer when we study the particular cases.

Let us consider the 5-dimensional case first.

\subsection{T~duality of the 5-dimensional solutions}
\label{sec-Tduality5}

We are going to start by assuming that there is a choice of charge parameters
whose T~duality transformations do not involve $\alpha'$ corrections. In the
case at hands, these are the hatted charge parameters $\hat{q}_{\pm,0}$ and
they are such that the $\mathcal{Z}_{\pm,0}$ functions can be written in the
form\footnote{The corrections $\delta \mathcal{Z}_{\pm,0}$ do not necessarily
  coincide with the ones we have directly obtained. If the hatted parameters
  are different from the unhatted ones, the redefinition modifies the
  corrections $\delta \mathcal{Z}_{\pm,0}$.}

\begin{equation}
  \mathcal{Z}_{\pm,0}
  = 1+\frac{\hat{q}_{\pm,0}}{\rho^{2}} +\alpha'\delta\mathcal{Z}_{\pm,0}
  \equiv
  \mathcal{Z}_{\pm,0}^{(0)}+\alpha'\delta\mathcal{Z}_{\pm,0}\,.
\end{equation}

Substituting this expression into the 5-dimensional form of the ansatz
Eqs.~(\ref{eq:5dsolution}) and in Eqs.~(\ref{eq:k15d}) and using the modified-Einstein-frame metric (\ref{eq:5dmetric}), we find

\begin{subequations}
\label{eq:5dsolutionwithcorrections}
\begin{align}
ds_{E}^{2}
  & =
    f^{(0)\, 2}\left[1
    -\frac{2\alpha'}{3}\left(\frac{\delta\mathcal{Z}_{+}}{\mathcal{Z}^{(0)}_{+}}
    +\frac{\delta\mathcal{Z}_{-}}{\mathcal{Z}^{(0)}_{-}}
    +\frac{\delta\mathcal{Z}_{0}}{\mathcal{Z}^{(0)}_{0}}\right)\right]dt^{2}
    \nonumber \\
  & \nonumber \\
  & \hspace{.5cm}
    -f^{(0)\, -1}\left[1
    +\frac{\alpha'}{3}\left(\frac{\delta\mathcal{Z}_{+}}{\mathcal{Z}^{(0)}_{+}}
    +\frac{\delta\mathcal{Z}_{-}}{\mathcal{Z}^{(0)}_{-}}
    +\frac{\delta\mathcal{Z}_{0}}{\mathcal{Z}^{(0)}_{0}}\right)\right]
    \left(d\rho^{2}+\rho^{2}d\Omega_{(3)}^{2}\right)\,,
  \\
& \nonumber \\
H^{(1)}
& = 
-\beta_{0}\left(\hat{q}_{0}-\alpha'\rho^{3}\delta\mathcal{Z}_{0}'\right)\omega_{(3)}\,,
\\
& \nonumber \\
  A
  & =
   \beta_{+}k_{\infty}^{-1}
    \left[-1+\frac{1}{\mathcal{Z}^{(0)}_{+}}\left(1-\alpha'\frac{\delta\mathcal{Z}_{+}}{\mathcal{Z}^{(0)}_{+}}\right)\right]dt \,,
  \\
& \nonumber \\
  B^{(1)}
  & =
    \beta_{-}k_{\infty}\left\{-1 +\frac{1}{\mathcal{Z}^{(0)}_{-}}
    \left[1
    +\alpha'\left(\Delta^{(0)}-\frac{\delta\mathcal{Z}_{-}}{\mathcal{Z}^{(0)}_{-}}\right)
    \right]\right\}dt\,,
  \\
& \nonumber \\
e^{-2\phi}
  & =
    e^{-2\phi_{\infty}}
   \frac{\sqrt{\mathcal{Z}^{(0)}_{+}\mathcal{Z}^{(0)}_{-}}}{\mathcal{Z}^{(0)}_{0}}
    \left\{
    1+\frac{\alpha'}{2}\left(\frac{\delta\mathcal{Z}_{+}}{\mathcal{Z}^{(0)}_{+}}+\frac{\delta\mathcal{Z}_{-}}{\mathcal{Z}^{(0)}_{-}}-2\frac{\delta\mathcal{Z}_{0}}{\mathcal{Z}^{(0)}_{0}}\right)
    \right\}\,,
\\
& \nonumber \\
  k
  & =
  k_{\infty}\sqrt{\frac{\mathcal{Z}^{(0)}_{+}}{\mathcal{Z}^{(0)}_{-}}}
  \left\{
    1+\frac{\alpha'}{2}\left(\frac{\delta\mathcal{Z}_{+}}{\mathcal{Z}^{(0)}_{+}}-\frac{\delta\mathcal{Z}_{-}}{\mathcal{Z}^{(0)}_{-}}\right)
    \right\}\,,    
\\
& \nonumber \\
  k^{(1)}
  & =
  k_{\infty}\sqrt{\frac{\mathcal{Z}^{(0)}_{+}}{\mathcal{Z}^{(0)}_{-}}}
  \left\{
    1+\frac{\alpha'}{2}\left(\frac{\delta\mathcal{Z}_{+}}{\mathcal{Z}^{(0)}_{+}}-\frac{\delta\mathcal{Z}_{-}}{\mathcal{Z}^{(0)}_{-}} +2\Delta^{(0)}\right)
    \right\}\,.    
\end{align}
\end{subequations}

\noindent
where

\begin{subequations}
  \begin{align}
    f^{(0)\,-1}
    & =
      \left(\mathcal{Z}^{(0)}_{+}\mathcal{Z}^{(0)}_{-}\mathcal{Z}^{(0)}_{0}\right)^{1/3}
      =
      \frac{1}{\rho^{2}}\left[(\rho^{2}+\hat{q}_{+})(\rho^{2}+\hat{q}_{-})(\rho^{2}+\hat{q}_{0})\right]^{1/3}\,,
    \\
    & \nonumber \\
    \Delta^{(0)} & =
  \tfrac{1}{4}(1+\beta_{+}\beta_{-})
 \frac{\mathcal{Z}^{(0)\, \prime}_{+}\mathcal{Z}^{(0)\,
                   \prime}_{-}}{\mathcal{Z}^{(0)}_{0}\mathcal{Z}^{(0)}_{+}\mathcal{Z}^{(0)}_{-}}
                   =
\frac{(1+\beta_{+}\beta_{-})\hat{q}_{+}\hat{q}_{-}}{(\rho^{2}+\hat{q}_{+})(\rho^{2}+\hat{q}_{-})(\rho^{2}+\hat{q}_{0})}\,.
  \end{align}
\end{subequations}

T~duality along the internal $z$ direction interchanges the 5-dimensional KK
and winding vectors $A$ and $B^{(1)}$ and transforms $k$ into $k^{(1)\, -1}$,
leaving the rest of the fields invariant \cite{Elgood:2020xwu}. The invariance
of the ansatz under these transformations implies that they are equivalent to
the transformations of the parameters in
Eq.~(\ref{eq:Tdualitytransformationoftheparametersd=5}). The functions
$f^{(0)}$ and $\Delta^{(0)}$ are invariant under these transformations. As for
the corrections $\delta\mathcal{Z}_{\pm,0}$, they must transform according to

\begin{equation}
  \delta\mathcal{Z}_{\pm}'
  =
  \delta\mathcal{Z}_{\mp} \mp\Delta^{(0)}\mathcal{Z}^{(0)}_{\mp}\,,
\end{equation}

\noindent
while $\delta\mathcal{Z}_{0}$ remains invariant, in agreement with
Eq.~(\ref{eq:Z0correctedd5}). An equivalent expression of the same result is
that the combination

\begin{equation}
  \frac{\delta\mathcal{Z}_{+}}{\mathcal{Z}_{+}^{(0)}}
  -\frac{\delta\mathcal{Z}_{-}}{\mathcal{Z}_{-}^{(0)}}
  +\Delta^{(0)}\,,
\end{equation}

\noindent
must be odd under T~duality.

Then, if we find that $\delta\mathcal{Z}_{-}=0$, we automatically know
$\delta\mathcal{Z}_{+}$, which is given by

\begin{equation}
  \delta\mathcal{Z}_{+}
  =
 \Delta^{(0)}\mathcal{Z}^{(0)}_{-}\,,
\end{equation}

\noindent
in agreement with Eq.~(\ref{eq:Z+correctedd5}).

\subsection{T~duality of the 4-dimensional solutions}
\label{sec-Tduality4}

Under the same assumptions made in the 5-dimensional case for the functions
$\mathcal{Z}_{\pm,0},\mathcal{H}$,\footnote{Again, the corrections
  $\delta \mathcal{Z}_{\pm,0}$ are not necessarily the ones we have defined
  originally.}  and expanding the constant $C_{\phi}$ as

\begin{equation}
  C_{\phi}
  =
  e^{2\phi_{\infty}}\left(1+\alpha'\delta C\right)\,,
\end{equation}

\noindent
we can rewrite the non-vanishing fields of the ansatz for the 4-dimensional solutions in
Eqs.~(\ref{eq:4dimensionalansatz}) as

\begin{subequations}
  \begin{align}
    ds_{E}^{2}
    & =
      e^{2U^{(0)}}\left[1-\frac{\alpha'}{2}\left(-2\delta C -2\frac{\delta q_{-}}{q_{-}}
      +2\frac{r^{2}\delta\mathcal{Z}_{-}'}{\hat{q}_{-}}+\frac{\delta\mathcal{Z}_{+}}{\mathcal{Z}^{(0)}_{+}}+\frac{\delta\mathcal{Z}_{-}}{\mathcal{Z}^{(0)}_{-}}+\frac{\delta\mathcal{Z}_{0}}{\mathcal{Z}^{(0)}_{0}}+\frac{\delta\mathcal{H}}{\mathcal{H}^{(0)}}\right)
   \right] dt^{2}
      \nonumber \\
    & \nonumber \\
    & \hspace{.5cm}
      -e^{2U^{(0)}}\left[1+\frac{\alpha'}{2}\left(2\delta C +2\frac{\delta q_{-}}{q_{-}}
      -2\frac{r^{2}\delta\mathcal{Z}_{-}'}{\hat{q}_{-}}+\frac{\delta\mathcal{Z}_{+}}{\mathcal{Z}^{(0)}_{+}}+\frac{\delta\mathcal{Z}_{-}}{\mathcal{Z}^{(0)}_{-}}+\frac{\delta\mathcal{Z}_{0}}{\mathcal{Z}^{(0)}_{0}}+\frac{\delta\mathcal{H}}{\mathcal{H}^{(0)}}\right)
      \right] \times
      \nonumber \\
    & \nonumber \\
    & \hspace{.5cm}
     \times \left(dr^{2}+r^{2}d\Omega^{2}_{(2)}\right)\,,
    \\
   & \nonumber \\
   A^{w}
   & =
       \beta q\ell^{-1}_{\infty}\cos{\theta}d\varphi\,,
    \\
  &  \nonumber \\
   A^{z}
   & =
       \beta_{+}k^{-1}_{\infty}\left[-1+\frac{1}{\mathcal{Z}^{(0)}_{+}}
       \left(1-\alpha'\frac{\delta\mathcal{Z}_{+}}{\mathcal{Z}^{(0)}_{+}}\right)\right]dt\,,
    \\
  &  \nonumber \\
  B^{(1)}{}_{w}
    & =
    \beta_{0}\ell_{\infty}\hat{q}_{0h}\cos{\theta}d\varphi\,,
    \\
    & \nonumber \\
  B^{(1)}{}_{z}
    & =
    \beta_{-}k_{\infty}\left\{-1 +\frac{1}{\mathcal{Z}^{(0)}_{-}}
      \left[1+\alpha'\left(\Delta^{(0)}-\frac{\delta\mathcal{Z}_{-}}{\mathcal{Z}^{(0)}_{-}}
      \right)\right]\right\}  dt\,,
    \\
    & \nonumber \\
    \ell
    & =
      \ell_{\infty}\sqrt{\frac{\mathcal{Z}^{(0)}_{0}}{\mathcal{H}^{(0)}}}\
      \left[1+\frac{\alpha'}{2}\left(\frac{\delta\mathcal{Z}_{0}}{\mathcal{Z}^{(0)}_{0}}
      -\frac{\delta\mathcal{H}}{\mathcal{H}^{(0)}}\right)\right]\,,
    \\
    & \nonumber \\
    \ell^{(1)}
    & =
      \ell_{\infty}\sqrt{\frac{\mathcal{Z}^{(0)}_{0}}{\mathcal{H}^{(0)}}}\
      \left[1+\frac{\alpha'}{2}\left(\frac{\delta\mathcal{Z}_{0}}{\mathcal{Z}^{(0)}_{0}}
      -\frac{\delta\mathcal{H}}{\mathcal{H}^{(0)}}+2\Sigma^{(0)}\right)\right]\,,
    \\
    & \nonumber \\
    k
    & =
        k_{\infty}\sqrt{\frac{\mathcal{Z}^{(0)}_{+}}{\mathcal{Z}^{(0)}_{-}}}
  \left\{
    1+\frac{\alpha'}{2}\left(\frac{\delta\mathcal{Z}_{+}}{\mathcal{Z}^{(0)}_{+}}-\frac{\delta\mathcal{Z}_{-}}{\mathcal{Z}^{(0)}_{-}}\right)
    \right\}\,,    
\\
& \nonumber \\
  k^{(1)}
  & =
  k_{\infty}\sqrt{\frac{\mathcal{Z}^{(0)}_{+}}{\mathcal{Z}^{(0)}_{-}}}
  \left[
    1+\frac{\alpha'}{2}\left(\frac{\delta\mathcal{Z}_{+}}{\mathcal{Z}^{(0)}_{+}}-\frac{\delta\mathcal{Z}_{-}}{\mathcal{Z}^{(0)}_{-}} +2\Delta^{(0)}\right)
    \right]\,,
    \\
    & \nonumber \\
      e^{2\phi}
    & =
      e^{2\phi^{(0)}}
 \left[
      1+\frac{\alpha'}{2}\left(
      2\delta C +2\frac{\delta q_{-}}{q_{-}}
      -2\frac{r^{2}\delta\mathcal{Z}_{-}'}{\hat{q}_{-}}
      +\frac{\delta\mathcal{Z}_{0}}{\mathcal{Z}^{(0)}_{0}}
      +\frac{\delta\mathcal{H}}{\mathcal{H}^{(0)}}
      -\frac{\delta\mathcal{Z}_{+}}{\mathcal{Z}^{(0)}_{+}}
      -\frac{\delta\mathcal{Z}_{-}}{\mathcal{Z}^{(0)}_{-}}\right)
    \right]\,,
  \end{align}
\end{subequations}

\noindent
where we have defined

\begin{subequations}
  \begin{align}
    e^{2U^{(0)}}
    & =
      \left(\mathcal{Z}^{(0)}_{+}\mathcal{Z}^{(0)}_{-}\mathcal{Z}^{(0)}_{0}\mathcal{H}^{(0)}\right)^{1/2}
      =
      \frac{1}{r^{2}}
      \left[(r+\hat{q}_{+})(r+\hat{q}_{-})(r+\hat{q}_{0})(r+\hat{q})\right]\,,
    \\
    & \nonumber \\
    \Delta^{(0)}
    & =
      \tfrac{1}{4}(1+\beta_{+}\beta_{-})
      \frac{\hat{q}_{+}\hat{q}_{-}}{(r+\hat{q}_{+})(r+\hat{q}_{-})(r+\hat{q}_{0})(r+\hat{q})}\,,
    \\
    & \nonumber \\
    \Sigma^{(0)}
    & =
    -\frac{1}{4(\mathcal{Z}^{(0)}_{0}\mathcal{H}^{(0)})^{3}}
      \left[(\mathcal{Z}^{(0)\, \prime}_{0}\mathcal{H}^{(0)})^{2}
      +(\mathcal{Z}^{(0)}_{0}\mathcal{H}^{(0)\, \prime})^{2}
      +(\beta\beta_{0}-1)
      \mathcal{Z}^{(0)\, \prime}_{0}\mathcal{H}^{(0)\, \prime}
      \mathcal{Z}^{(0)}_{0}\mathcal{H}^{(0)}
      \right]
      \nonumber \\
    & \nonumber \\
    & =
    -\frac{1}{4(r+\hat{q}_{0})^{3}(r+\hat{q})^{3}}
      \left[\hat{q}_{0}^{2}(r+\hat{q})^{2}
      +\hat{q}^{2}(r+\hat{q}_{0})^{2}
      +(\beta\beta_{0}-1)\hat{q}_{0}\hat{q} (r+\hat{q}_{0})(r+\hat{q})
      \right]\,,
    \\
    & \nonumber \\
    e^{2\phi^{(0)}}
    & =
e^{2\phi_{\infty}}
      \sqrt{\frac{\mathcal{Z}^{(0)}_{0}\mathcal{H}^{(0)}}{\mathcal{Z}^{(0)}_{+}\mathcal{Z}^{(0)}_{-}}}
      =
e^{2\phi_{\infty}}
\sqrt{\frac{(r+\hat{q}_{0})(r+\hat{q})}{(r+\hat{q}_{+})(r+\hat{q}_{-})}}\,.
  \end{align}
\end{subequations}

Observe that, since, in general, $\mathcal{Z}_{0}\neq \mathcal{Z}_{0h}$, we
have to distinguish between $\hat{q}_{0}$ and $\hat{q}_{0h}$. The latter only
occurs in $B^{(1)}{}_{w}$.

Furthermore, observe that the asymptotic behaviour of the metric and dilaton
field are correct only if\footnote{In the solutions that we have found,
  $\delta C=\delta q_{-}=0$ for the $\beta_{0}=\beta$ one and
  $\delta C= -\frac{\delta q_{-}}{q_{-}}=\tfrac{1}{4}Y$.}

\begin{equation}
\delta C +\frac{\delta q_{-}}{q_{-}}=0\,.  
\end{equation}

The solutions described by this ansatz are expected to be invariant under
T~duality transformations in the internal circles parametrized by the
coordinates $w$ and $z$ ( $T_{w}$ and $T_{z}$). Let us start with $T_{w}$.

Invariance under $T_{w}$ interchanges $A^{w}$ and $B^{(1)}{}_{w}$ and $\ell$
and $\ell^{(1)\, -1}$, leaving the rest of the fields invariant. The
interchange between $A^{w}$ and $B^{(1)}{}_{w}$ implies that $\hat{q}=q$ and
the following transformation 

\begin{equation}
\beta \rightleftharpoons \beta_{0}\,
\hspace{.5cm}
\hat{q}  \rightleftharpoons \hat{q}_{0h}\,,
\hspace{.5cm}
\ell_{\infty}  \rightleftharpoons \ell_{\infty}^{(1)\, -1}\,.
\end{equation}

The interchange of $\ell$ and $\ell^{(1)\, -1}$ demands, at the level of
charge parameters, the interchange between $\hat{q}$ and $\hat{q}_{0}$ which,
in its turn, implies $\hat{q}_{0h}=\hat{q}_{0}$. Furthermore, the
transformations of the corrections $\delta\mathcal{Z}_{0}$ and
$\delta\mathcal{H}$ must be such that the combination

\begin{equation}
  \label{eq:aquellacondicion}
\frac{\delta\mathcal{Z}_{0}}{\mathcal{Z}^{(0)}_{0}}
-\frac{\delta\mathcal{H}}{\mathcal{H}^{(0)}}
+\Sigma^{(0)}\,,
\end{equation}

\noindent
is odd under $T_{w}$.

The invariance of $A^{z}$ and $B_{z}$ leads to the invariance of the
corrections $\delta\mathcal{Z}_{\pm}$ under $T_{w}$:

\begin{equation}
\delta\mathcal{Z}_{\pm}'  =\delta\mathcal{Z}_{\pm}\,.
\end{equation}

\noindent
This relation makes $k$ automatically invariant under $T_{w}$ as well.

Finally, taking into account the previous results, the invariance of the
dilaton and of the metric function lead to the invariance of the combination

\begin{equation}
\frac{\delta\mathcal{Z}_{0}}{\mathcal{Z}^{(0)}_{0}}
+\frac{\delta\mathcal{H}}{\mathcal{H}^{(0)}}\,.
\end{equation}

\noindent
Combining this condition with Eq.~(\ref{eq:aquellacondicion}) we find

\begin{subequations}
  \begin{align}
    \delta \mathcal{Z}_{0}'
    & = \delta\mathcal{H} -2\Sigma^{(0)}\mathcal{H}\,,
    \\
    & \nonumber \\
    \delta \mathcal{H}'
    & = \delta \mathcal{Z}_{0} +2\Sigma^{(0)}\mathcal{Z}_{0}\,.
  \end{align}
\end{subequations}

It can be checked that these relations are satisfied by the corrections we
have found. If we use in them additional information such as the vanishing of
$\delta\mathcal{H}$, these relations can be used to determine some of the
corrections in terms of $\Sigma^{(0)}$.

Let us now consider the effect of $T_{z}$. This transformation interchanges
$A^{z}$ and $B^{(1)}{}_{z}$, which implies the following transformation on the
parameters of the solution:

\begin{equation}
  \hat{q}_{\pm}\rightleftharpoons \hat{q}_{\mp}\,,
  \hspace{1cm}
  \beta_{\pm}\rightleftharpoons \beta_{\mp}\,,
  \hspace{1cm}
  k_{\infty} \rightleftharpoons k_{\infty}^{-1}\,,
\end{equation}

\noindent
and the same relations between the corrections $\delta\mathcal{Z}_{\pm}$ that
we obtained in the 5-dimensional case\footnote{Notice that $\Delta^{(0)}$ is
  different, though.}

\begin{equation}
  \delta\mathcal{Z}_{\pm}'
  =
  \delta\mathcal{Z}_{\mp} \mp\Delta^{(0)}\mathcal{Z}^{(0)}_{\mp}\,,
\end{equation}

These relations imply the relation we would obtain from the transformation
of the KK scalar $k$ and allows the determination
of $\delta\mathcal{Z}_{+}$ if we know that $\delta\mathcal{Z}_{-}=0$, for
instance. Furthermore, it implies the invariance of the combination

\begin{equation}
\frac{\delta\mathcal{Z}_{+}}{\mathcal{Z}_{+}^{(0)}}  
+\frac{\delta\mathcal{Z}_{-}}{\mathcal{Z}_{-}^{(0)}}\,,  
\end{equation}

\noindent
which occurs in the dilaton and metric function. The invariance of these two
fields requires

\begin{equation}
\frac{\delta\mathcal{Z}_{0}'}{\mathcal{Z}_{0}^{(0)}}  
+\frac{\delta\mathcal{H}'}{\mathcal{H}^{(0)}}  
=
\frac{\delta\mathcal{Z}_{0}}{\mathcal{Z}_{}^{(0)}}  
+\frac{\delta\mathcal{H}}{\mathcal{H}^{(0)}}
-2r^{2}\frac{d~}{dr}\left[\frac{\delta\mathcal{Z}_{-}}{q_{-}}
  -\frac{\delta\mathcal{Z}_{+}}{q_{+}}
  -\Delta^{(0)}\frac{\mathcal{Z}_{+}}{q_{+}}\right]\,,
\end{equation}

\noindent
while the invariance of the KK scalar $\ell$ requires that of

\begin{equation}
\frac{\delta\mathcal{Z}_{0}}{\mathcal{Z}_{0}^{(0)}}  
-\frac{\delta\mathcal{H}}{\mathcal{H}^{(0)}}\,.  
\end{equation}
Combining these two conditions, we get conditions for $\delta\mathcal{Z}_{0}$
and $\delta\mathcal{H}$ separately

\begin{subequations}
  \begin{align}
\delta\mathcal{Z}_{0}'
=
\delta\mathcal{Z}_{0}
-r^{2}\frac{d~}{dr}\left[\frac{\delta\mathcal{Z}_{-}}{q_{-}}
  -\frac{\delta\mathcal{Z}_{+}}{q_{+}}
    -\Delta^{(0)}\frac{\mathcal{Z}_{+}}{q_{+}}\right]\mathcal{Z}_{0}^{(0)}\,,
    \\
    & \nonumber \\
\delta\mathcal{H}'
=
\delta\mathcal{H}
-r^{2}\frac{d~}{dr}\left[\frac{\delta\mathcal{Z}_{-}}{q_{-}}
  -\frac{\delta\mathcal{Z}_{+}}{q_{+}}
    -\Delta^{(0)}\frac{\mathcal{Z}_{+}}{q_{+}}\right]\mathcal{H}^{(0)}\,.
  \end{align}
\end{subequations}
In the solutions we have found, these conditions are satisfied in an almost
trivial fashion, since the term in the derivative vanishes identically and
$\delta\mathcal{Z}_{0}$ and $\delta\mathcal{H}$ are invariant under $T_{z}$.

\section{Discussion}
\label{sec-discussion}

We have computed the $\alpha'$ corrections to a general family of 3- and 4-
charge black hole solutions of the heterotic superstring effective
action. These corrections are more easily obtained in ten dimensions, where
the solutions can be interpreted as a superposition of fundamental strings,
momentum waves, solitonic 5-branes, and (in the 4-charge case) Kaluza-Klein
monopoles.\footnote{This interpretation is correct at least in the
  supersymmetric case.} We have then derived the form of the 4- and
5-dimensional fields by using the results of Ref.~\cite{Ortin:2020xdm}. As we
have shown, a particular choice of the signs of the charges corresponds to
supersymmetric solutions, in which case we recover the results of
Refs.~\cite{Cano:2018qev,Cano:2018brq}. However, any other choice does not
preserve any supersymmetry. Interestingly, while at zeroth order in $\alpha'$
all these solutions have the same form, the first-order corrections are
drastically different depending on the way we break supersymmetry. These
differences not only affect the profile of the fields, but also the values of
the mass, area, entropy and charges of these extremal black holes. In fact, as
we have argued in the text, the identification of the physically relevant
charges becomes a non-trivial problem in the presence of $\alpha'$ terms. As a
practical solution, we have introduced the charge parameters
$\hat{\mathcal{Q}}_i$ as one would define them in the zeroth-order theory (see
Eqs.~(\ref{eq:5dchargesdef}) and (\ref{eq:charges4dcase1})). These charges
correspond to the leading coefficients in the asymptotic expansion of the
gauge fields in four or five dimensions and they would in principle correspond
to Maxwell charges as introduced in \cite{Marolf:2000cb}. Moreover, it has been
checked \cite{Faedo:2019xii, Cano:2021dyy} that these charges match those used
in the computation of near-horizon geometries
\cite{Sahoo:2006pm,Sen:2007qy,Prester:2008iu}, at least in the supersymmetric
cases. Thus, they serve us to make contact with previous results. There are,
nevertheless, other possible definitions of charges due to the presence of
Chen-Simons terms, and it is not clear which ones should be quantized. All
these issues clearly deserve further attention.

One of the most important results of this paper is the computation of the
black hole entropy in terms of these asymptotic charges. As explained in the
introduction, obtaining the $\alpha'$-corrected entropy in heterotic string
theory has been traditionally problematic due to the Lorentz-Chern-Simons
terms in the definition of 3-form field-strength $H$. Here we have performed a
rigorous computation of the entropy by using the gauge-invariant
generalization of Wald's formula provided in Ref.~\cite{Elgood:2020nls}. The
result, that we repeat here for convenience, reads

\begin{equation}
S=2\pi \sqrt{|\hat{\mathcal{Q}}_{+}\hat{\mathcal{Q}}_{-}|\left(k+2+\beta_{+}\beta_{-}\right)}\, ,
\end{equation}
where

\begin{equation}
k=\begin{cases}
|\hat{\mathcal{Q}}_{0}| &\text{if}\,\, D=5\\
|\hat{\mathcal{Q}}_{0}\hat{\mathcal{Q}}| +\beta_{0}\beta &\text{if}\,\,  D=4\, ,
\end{cases}
\end{equation}

\noindent
and where $\beta_{i}=\operatorname{sign}{(\hat{\mathcal{Q}}_{i})}$ is the sign
of the corresponding charge. The $D=5$ result matches the one in
Ref.~\cite{Prester:2008iu} using the entropy function method, provided one
relates $|\hat{\mathcal{Q}}_{0}|$ with the number of S5-branes
correctly.\footnote{Ref.~\cite{Prester:2008iu} expresses the five-dimensional
  result in terms of the number of S5-branes $N$ obtained by integrating the
  brane-source charge, and in fact we have $|\hat{\mathcal{Q}}_{0}| =N-1$, 
  see \cite{Faedo:2019xii,Cano:2018hut,Cano:2021dyy}.} The supersymmetric case
also matches the corrections to the statistical entropy obtained in
Ref.~\cite{Castro:2008ys}. The $D=4$ result in the supersymmetric case
$\beta_0\beta=\beta_{+}\beta_{-}=1$ matches exactly previous microscopic
results \cite{Kutasov:1998zh, Kraus:2005vz, Sen:2007qy}. It also matches the
results of refs.~\cite{Sahoo:2006pm, Sen:2007qy,Prester:2008iu, Faedo:2019xii,
  Cano:2021dyy} that applied the entropy function method for the cases
$\beta_0\beta=1$ and arbitrary $\beta_{+}\beta_{-}$. To the best of our
knowledge, the case $\beta_0\beta=-1$ is new.  Although this entropy has been
computed from the first-order in $\alpha'$ action, it can be argued that it is
exact in the $\alpha'$ expansion \cite{Kraus:2005vz, Kraus:2006wn,
  Sen:2007qy}. As a matter of fact, the value of the entropy can be determined
from the near-horizon geometry only, and for all the black holes considered
here, this is an AdS$_3\times\mathbb{S}^3/\mathbb{Z}_n\times\mathrm{T}^4$
geometry with the peculiarity that the curvature of the torsionful spin
connection vanishes, $R_{(-)}=0$.  Hence, any higher-order term in the action
that contains more than one factor of $R_{(-)}$ will be irrelevant for the
entropy. This is certainly the case for the $\mathcal{O}(\alpha'^3)$ terms
introduced by supersymmetry \cite{Bergshoeff:1989de}.  It cannot be guaranteed
that the effective action does not contain other types of terms, but this
argument at least protects the entropy from supersymmetry-related
higher-derivative corrections.

It is also interesting to look at the mass of these solutions, which receives
$\alpha'$ corrections.  We can analyze this from the viewpoint of the Weak
Gravity Conjecture (WGC), according to which, the corrections to the
extremality bound should be such that the decay of an extremal black hole into
a set of smaller black holes should be possible in terms of charge and energy
conservation \cite{Kats:2006xp,Cheung:2018cwt,Hamada:2018dde}.  This has
mostly been applied to theories with a single $\mathrm{U}(1)$ field
\cite{Aalsma:2019ryi,Bellazzini:2019xts,Cremonini:2019wdk,Charles:2019qqt,Cano:2019ycn,Loges:2019jzs,Andriolo:2020lul,Loges:2020trf,Cano:2020qhy,Cano:2021tfs,Arkani-Hamed:2021ajd},
in which case this is rephrased as the condition that the charge-to-mass ratio
of extremal black holes is corrected positively.  For dimensionality reasons,
four-derivative corrections always modify the charge-to-mass ratio as
$Q/M=1+c/Q^2+\ldots$, for a certain constant $c$. However, when there are
several charges involved, the corrections to the extremal mass can take a more
complicated form. Also, in that case one needs to formulate the WGC
appropriately \cite{Jones:2019nev}.  For our black holes we have found that,
in most instances, the mass is not corrected when expressed in terms of the
asymptotic charges $\hat{\mathcal{Q}}_i$. This is the case in particular for
supersymmetric black holes. However, the non-supersymmetric solution with
$\beta_{0}\beta=-1$ does contain an interesting correction to the mass, given
by Eq.~(\ref{eq:shiftM4d}). This depends non-trivially on the charges
$\hat{\mathcal{Q}}_{0}$ and $\hat{\mathcal{Q}}$, and we have checked that it
is negative for any of their values. Having a negative shift to the mass seems
to be the natural generalization of the WGC for this class of black holes, so
this result provides a non-trivial check of the validity of the WGC in string
theory.

The zeroth-order solutions we have started from, as families of solutions, are
invariant under T~duality. We have seen how this property is preserved at
first order when the first-order T-duality rules of
Ref.~\cite{Bergshoeff:1995cg} are used. As a matter of fact, the
lower-dimensional solutions can be written in a manifestly T-duality-invariant
form using the relations between the higher- and lower-dimensional fields of
Ref.~\cite{Elgood:2020xwu}. A fundamental property of those relations is the
presence of $\alpha'$ corrections in them for all the fields descending from
the 10-dimensional Kalb-Ramond 2-form. Since these corrections depend only on
the zeroth-order fields (at first order) they, automatically, determine part
of the $\alpha'$ corrections to the solutions and, as we have discussed in
Section~\ref{sec-Tduality}, this fact and T~duality can be used to determine
some of the first-order $\alpha'$ corrections to all the fields.

Finally, we find it worth remarking that, interestingly, some of the solutions
that break supersymmetry are simpler than their supersymmetric counterparts. We are
referring to the case $\beta_{+}\beta_{-}=-1$, and $\beta_{0}\beta=1$. For
that choice of signs, only the function $\mathcal{Z}_0$ receives
corrections. In addition, we know that those corrections can be cancelled by
turning on the non-Abelian gauge fields and introducing one (in $D=5$) or two (in
$D=4$) $\mathrm{SU}(2)$ instantons
\cite{Cano:2018qev,Chimento:2018kop,Cano:2018brq}. In that situation, the
zeroth-order solution receives no $\alpha'$ corrections whatsoever, (except
for the presence of the instantons), and this is because the $T$-tensors
defined in Eq.~(\ref{eq:Ttensors}), and that control the $\alpha'$
corrections, are identically vanishing. Interestingly, the next-order
corrections appear at $\mathcal{O}(\alpha'^3)$ and they are again built out of
these $T$-tensors \cite{Bergshoeff:1989de}, so they do not modify these solutions
either. One could speculate that the full tower of $\alpha'$ terms is built
out of these tensors, in which case these black holes would be $\alpha'$-exact
solutions of the Heterotic Superstring effective action. This may actually be
the case for the higher-derivative terms that arise from the
supersymmetrization of the Lorentz-Chern-Simons terms, which are the ones
computed in \cite{Bergshoeff:1989de}. However, the effective action could
contain other contributions besides those, which would invalidate the claim on
the exactness of these solutions. Unfortunately, our current knowledge on the
higher-order $\alpha'$ terms is too limited to establish a definitive
conclusion.

\section*{Acknowledgments}

This work has been supported in part by the MCIU, AEI, FEDER (UE) grant
PGC2018-095205-B-I00 and by the Spanish Research Agency (Agencia Estatal de
Investigaci\'on) through the grant IFT Centro de Excelencia Severo Ochoa
SEV-2016-0597.  The work of PAC is supported by a postdoctoral fellowship from
the Research Foundation - Flanders (FWO grant 12ZH121N). The work of AR is
supported by a postdoctoral fellowship associated to the MIUR-PRIN contract
2017CC72MK003. The project that gave rise to these results received the
support of a fellowship from ``la Caixa'' Foundation (ID 100010434). The
fellowship code is LCF/BQ/DI20/11780035.  TO wishes to thank M.M.~Fern\'andez
for her permanent support.

\appendix

\section{Relation between 10- and 9-dimensional fields at $\mathcal{O}(\alpha')$}
\label{app-dictionaryfirstorder}

At first order in $\alpha'$, the 10-dimensional fields can be expressed in
terms of the 9-dimensional ones as follows:

\begin{equation}
  \begin{aligned}
    \hat{g}_{\mu\nu} & =
    g_{\mu\nu} -k^{2}A_{\mu}A_{\nu}\,,
    \\
    & \\
    \hat{g}_{\mu \underline{z}}  & =
    -k^{2}A_{\mu}\,,
    \\
        & \\
    \hat{g}_{\underline{z}\underline{z}}
    & =
    -k^{2}\,,
    \\
        & \\
         \hat{B}_{\mu\nu} & =
     B^{(1)}{}_{\mu\nu}
  -A_{[\mu}
  \left[
    B^{(1)}{}_{\nu]}
    +\frac{\alpha'}{2}
    k\left(\varphi_{A}A^{A}{}_{|\nu]}
        +\tfrac{1}{2}\Omega^{(0)}_{(-)\, |\nu]}{}^{ab}K^{(+)}{}_{ab}
      -K^{(-)}{}_{|\nu]}{}^{a}\partial_{a}\log{k}\right)
    \right]\,,
    \\
        & \\
     \hat{B}_{\mu\underline{z}}
     & =
     B^{(1)}{}_{\mu}
      +\frac{\alpha'}{4}k\left(\varphi_{A}A^{A}{}_{\mu}
        +\tfrac{1}{2}\Omega^{(0)}_{(-)\, \mu}{}^{ab}K^{(+)}{}_{ab}
        -K^{(-)}{}_{\mu}{}^{a}\partial_{a}\log{k}\right)\,,
      \\
          & \\
    \hat{\phi}
    & =
    \phi+\tfrac{1}{2}\log{k}\,,
    \\
    & \\
    \hat{A}^{A}{}_{\mu}
    & =
    A^{A}{}_{\mu} +k\varphi^{A}A_{\mu}\,,
    \\
    & \\
    \hat{A}^{A}{}_{\underline{z}}
    & =
    k\varphi^{A}\,.
  \end{aligned}
\end{equation}

The inverse relations are

\begin{equation}
  \begin{aligned}
    g_{\mu\nu}
    & =
    \hat{g}_{\mu\nu}-\hat{g}_{\underline{z}\mu}
    \hat{g}_{\underline{z}\nu}/\hat{g}_{\underline{z}\underline{z}}\,,
    \\
    & \\
    A_{\mu}
    & =
    \hat{g}_{\mu \underline{z}}/\hat{g}_{\underline{z}\underline{z}}\,,
    \\
    & \\
    k
    & =
    |\hat{g}_{\underline{z}\underline{z}}|^{1/2}\,,
    \\
    & \\
    B^{(1)}{}_{\mu\nu}
    & =
    \hat{B}_{\mu\nu}
    +\hat{g}_{\underline{z}[\mu}
    \left[
         \hat{B}_{|\nu]\, \underline{z}}
      +\frac{\alpha'}{4}\left(\hat{A}^{A}{}_{|\nu]}\hat{A}_{A\, \underline{z}}
      +\hat{\Omega}^{(0)}_{(-)\, |\nu]}{}^{\hat{a}}{}_{\hat{b}}
      \hat{\Omega}^{(0)}_{(-)\, \underline{z}}{}^{\hat{b}}{}_{\hat{a}}
      \right)\right]
    /\hat{g}_{\underline{z}\underline{z}}\,,
    \\
    & \\
    B^{(1)}{}_{\mu}
    & =
     \hat{B}_{\mu\, \underline{z}}
      -\frac{\alpha'}{4}\left[\hat{A}^{A}{}_{\mu}\hat{A}_{A\, \underline{z}}
      +\hat{\Omega}^{(0)}_{(-)\, \mu}{}^{\hat{a}}{}_{\hat{b}}
      \hat{\Omega}^{(0)}_{(-)\, \underline{z}}{}^{\hat{b}}{}_{\hat{a}}
    \right.
    \\
    & \\
    &
    \left.
      -\hat{g}_{\mu \underline{z}}\left(\hat{A}^{A}{}_{\underline{z}}\hat{A}_{A\, \underline{z}}
      +\hat{\Omega}^{(0)}_{(-)\, \underline{z}}{}^{\hat{a}}{}_{\hat{b}}
      \hat{\Omega}^{(0)}_{(-)\, \underline{z}}{}^{\hat{b}}{}_{\hat{a}}\right)
  /\hat{g}_{\underline{z}\underline{z}}\right]\,,
    \\
    & \\
    \phi
    & =
    \hat{\phi}-\tfrac{1}{4}\log{(-\hat{g}_{\underline{z}\underline{z}})}\,,
    \\
    & \\
    A^{A}{}_{\mu}
    & =
    \hat{A}^{A}{}_{\mu} -\hat{A}^{A}{}_{\underline{z}}\hat{g}_{\mu\underline{z}}/\hat{g}_{\underline{z}\underline{z}}\,,
    \\
    & \\
    \varphi^{A}
    & =
    \hat{A}^{A}{}_{\underline{z}}/(-\hat{g}_{\underline{z}\underline{z}})^{1/2}\,.
  \end{aligned}
\end{equation}

We also use the scalar combination $k^{(1)}$, which is defined as follows:

\begin{equation}
  \label{eq:k1def}
  k^{(1)}
  \equiv
  k\left[1+\frac{\alpha'}{4}\left(\varphi^{2}-\tfrac{1}{4}K^{(+)\, 2}
      +2(\partial\log{k})^{2}\right)\right]\,.
\end{equation}

In terms of the 10-dimensional fields

\begin{equation}
  \label{eq:k1def2}
  k^{(1)}
  \equiv
  k\left[1-\frac{\alpha'}{4}\left(\hat{A}_{A\,\underline{z}}\hat{A}^{A}{}_{\underline{z}}+\hat{\Omega}^{(0)}_{(-)\,
      \underline{z}}{}^{\hat{a}}{}_{\hat{b}}\hat{\Omega}^{(0)}_{(-)\,
      \underline{z}}{}^{\hat{b}}{}_{\hat{a}}\right)/\hat{g}_{\underline{z}\underline{z}}\right]\,.
\end{equation}

Some of these expressions contain implicit $\mathcal{O}(\alpha'{}^{2})$ terms
which give them a nicer or simpler form. They should be consistently ignored.

\section{The coset space SU$(2)/$U$(1)$}
\label{app:coset}

The su$(2)$ algebra 

\begin{equation}
  [T_{i},T_{j}] = - \epsilon_{ijk}T_{k}\,,
  \hspace{1cm}
  i,j,k=1,2,3\,,
\end{equation}

\noindent
can be split into horizontal ($P_{a}$) and vertical ($M$) components
with the commutation relations

\begin{equation}
  \label{eq:su2algebrasplit}
  [P_{a},P_{b}] = - \epsilon_{ab}M\,,
  \hspace{1cm}
  [M,P_{a}] = - \epsilon_{ab}P_{b}\,,
  \hspace{1cm}
  a,b=1,2\,.
\end{equation}

A SU$(2)/$U$(1)$ corset representative, $u$ can be constructed by
exponentiation

\begin{equation}
u= e^{x^{1}P_{1}}  e^{x^{2}P_{2}}\,,
\end{equation}

\noindent
where $x^{a}$ are two coordinates unrelated to those of the main body
of this paper.

The left-invariant Maurer-Cartan 1-form is

\begin{equation}
  \label{eq:MC1-form}
  \begin{aligned}
    V
    & = -u^{-1}du
    \\
    & \\
    & = (-\cos{x^{2}}dx^{1})P_{1} +(-dx^{2})P_{2}
    +\sin{x^{2}}dx^{1}M
    \\
    & \\
    & \equiv
    v^{a}P_{a}+\vartheta M\,.
  \end{aligned}
\end{equation}

The horizontal components $v^{a}$ will be used as Zweibeins while the vertical
component $\vartheta$ will play the role of U$(1)$ connection.  The
Maurer-Cartan equations are

\begin{equation}
  dV-V\wedge V=0\,,
  \,\,\,\,\,
  \Rightarrow
  \,\,\,\,\,
  \left\{
    \begin{array}{rcl}
      dv^{a}& = &  -\epsilon^{ab}v^{b}\wedge \vartheta\,,
      \\
            & & \\
      d\vartheta
            & = &
-\tfrac{1}{2}\epsilon_{ab}v^{a}\wedge v^{b}\,.
    \end{array}
    \right.
\end{equation}

The coordinates $x^{a}$ are related to the standard $\theta,\varphi$ by

\begin{equation}
  x^{2} = \pi/2-\theta\,,
  \hspace{1cm}
  x^{1} =-\varphi\,,
\end{equation}

\noindent
and, in terms of these coordinates

\begin{equation}
  \label{eq:componentsMC1-form}
  v^{1} = \sin{\theta}d\varphi\,,
  \hspace{.5cm}
  v^{2} = d\theta\,,
  \hspace{.5cm}
  \vartheta = -\cos{\theta}d\varphi\,.
\end{equation}

Using the invariant metric $\delta_{ab}$, we get

\begin{subequations}
  \begin{align}
    ds^{2}
    & =
      \delta_{ab}v^{a}\otimes v^{b} = d\Omega^{2}{}_{(2)}\,,
    \\
    & \nonumber \\
    \omega_{(2)}
    & =
      -\tfrac{1}{2}\epsilon_{ab}v^{a}\wedge v^{b}
      =
      \sin{\theta}d\theta\wedge d\varphi\,,
    \\
    & \nonumber \\
    d\vartheta
    & =
       \omega_{(2)}\,.
  \end{align}
\end{subequations}


\end{document}